\documentclass[showpacs,floatfix,superscriptaddress]{revtex4}
\usepackage{graphicx}
\usepackage{multirow}
\usepackage{epsfig}
\usepackage{bm}
\usepackage[utf8]{inputenc}
\usepackage{amssymb}
\usepackage{float}
\usepackage{amsmath}
\usepackage{dcolumn}
\usepackage{latexsym}
\usepackage{subfig}
\usepackage{amsthm}
\usepackage{array}
\usepackage{framed}
\usepackage{cancel}
\usepackage[colorlinks]{hyperref}
\usepackage[usenames,dvipsnames]{color}
\hypersetup{
     breaklinks=true,
    pdfstartview={FitH},    
    colorlinks=true,       
    linkcolor=blue,          
    citecolor=red,        
    filecolor=magenta,      
    urlcolor=blue,           
    anchorcolor=green,      
    linktocpage=true
}

\def\doi{http://doi.org}

\begin{document}

\title{Accretion onto a Charged Black Hole in Consistent 4D Einstein-Gauss-Bonnet Gravity}

\author{Kourosh Nozari}
\email[]{knozari@umz.ac.ir (Corresponding Author)}
\author{Sara Saghafi}
\email[]{s.saghafi@umz.ac.ir}
\author{Mohammad Hassani}
\email[]{Mohammad.hassanei@gmail.com}
\affiliation{Department of Theoretical Physics, Faculty of Science, University of Mazandaran,\\
P. O. Box 47416-95447, Babolsar, Iran}

\begin{abstract}
In astrophysics, accretion is the process by which a massive object acquires matter. The infall leads to the extraction of gravitational energy. Accretion onto dark compact objects such as black holes, neutron stars, and white dwarfs is a crucial process in astrophysics as it turns gravitational energy into radiation. The accretion process is an effective technique to investigate the properties of other theories of gravity by examining the behavior of their solutions with compact objects. In this paper, we investigate the behavior of test particles around a charged four-dimensional Einstein–Gauss–Bonnet ($4D$ EGB) black hole in order to understand their innermost stable circular orbit (ISCO) and energy flux, differential luminosity, and temperature of the accretion disk. Then, we examine particle oscillations around a central object via applying restoring forces to treat perturbations. Next, we explore the accretion of perfect fluid onto a charged $4D$ EGB black hole. We develop analytical formulas for four-velocity and proper energy density of the accreting fluid. The EGB parameter and the charge affect properties of the test particles by decreasing their ISCO radius and also decreasing their energy flux.
Increasing the EGB parameter and the charge, near the central source reduces both the energy density and the radial component of the infalling fluid's four-velocity.
\vspace{12 pt}
\end{abstract}


\maketitle

\enlargethispage{\baselineskip}
\tableofcontents

\section{Introduction}\label{intro}

Einstein's theory of general relativity (GR) was proposed more than a century ago and has passed several observational tests, particularly in the weak field regime. One of the most striking results of GR is the ability to anticipate black holes. The existence of these physical objects is supported by observations of gravitational waves from black hole mergers by the LIGO/Virgo experiment \cite{Abbott_2016}, the M87* and SgA* supermassive black holes shadows by the Event Horizon Collaboration \cite{2019ApJ...875L...1E}, and the electromagnetic spectrum emitted by an accretion disk around a black hole \cite{Frank_2002_accretion,2014ARA&A..52..529Y,Nampalliwar_2020}. Nowadays, recent observations of black holes in gravitational and electromagnetic spectra, as well as future advances, have sparked much interest in testing GR and its alternatives in the strong gravity regime.\\

Observations of black holes in strong gravity regimes focus on a small area around the black hole, including a range from the photon sphere to the accretion disk. The significance of this region stems from the fact that it may be influenced by the possible higher curvature corrections to the Einstein term in GR. The Gauss-Bonnet term and its Lovelock generalization are the key higher curvature terms investigated in alternative theories beyond GR. In four dimensions, the Gauss-Bonnet term is a topological invariant that only affects gravitational dynamics when coupled to a matter field. Glavan and Lin \cite{Glavan_2020} suggested a new $4D$ EGB theory of gravity that incorporates a regularized Gauss-Bonnet term by rescaling the Gauss-Bonnet coupling constant  $\alpha  \to \frac{\alpha }{{D - 4}}$ in the limit $D \to 4$. This scaling reveals that the Gauss-Bonnet term contributes significantly to gravitational dynamics in the limit $D \to 4$. However, this alteration contradicts common knowledge and may cause complications. The theory is not clearly stated in the limit $D \to 4$ \cite{Ai_2020,PhysRevLett.125.149001,Gurses_2020,Li_2020,Kobayashi_2020,Hennigar_2020}. Furthermore, the model's vacua is ill-defined \cite{Aoki_2020}. To address these concerns, numerous variations of the original hypothesis have been investigated. In \cite{Gurses_2020}, it is demonstrated that the original $4D$ EGB theory can be reconstructed as a class of the Horndeski theory with an additional scalar degree of freedom. The Lovelock generalization as a scalar-tensor theory is also discussed in \cite{Li_2020}.
Similar conclusions have been investigated in \cite{Kobayashi_2020,Hennigar_2020} by adding a counterterm in $D$-dimensions and taking $D \to 4$ as the limit. Another regularization method involves breaking the theory's temporal diffeomorphism invariance \cite{Fernandes_2020_1}. A Lagrangian multiplier can eliminate the scalar degree of freedom, resulting in the theory having the same number of degrees of freedom as general relativity.\\

The original $4D$ EGB idea and its modifications have received significant attention recently. In this theoretical framework, there is a large amount of research devoted to understanding the nature of the new $4D$ EGB theory \cite{Aoki_2020,Liu_2021,Guo_2020_2,Wei_2021,Kumar_2020,Konoplya_2020_3,Churilova_2020_quasinormal,Malafarina_2020,Aragon_2020_perturbative,
Mansoori_2021,Ge_2020,Rayimbaev_2020_magnetized,Chakraborty_2020,Odintsov_2020_rectifying,Odintsov_2020,Lin_2020,
Shaymatov_2020,Islam_2020,Singh_2020}, the motion of spinning particles \cite{Zhang_2020}, and the scalar and electromagnetic perturbations in testing the strong cosmic censorship conjecture \cite{Mishra_2020}. The black hole solutions and their physical properties include shadows of charged black holes in AdS space \cite{Fernandes_2020_2}, and relativistic star solutions \cite{Doneva_2021}. Extensive analysis was conducted on stability of the Einstein static universe \cite{Li_2020_stability}, gravitational lensing \cite{Jin_2020}, speed of gravitational waves and scalar perturbations \cite{Feng_2021}, observational constraints on the $4D$ EGB theoretical parameters $\alpha$ \cite{Feng_2021,Clifton_2020}, and thermodynamic geometry and phase transitions \cite{Wei_2020,Hegde_2020_thermodynamics}. \\

Accretion is the process of dragging particles onto a black hole. This process transfers energy into the environment, resulting in astronomical phenomena like powerful jets, high-energy radiation, and quasars \cite{Kato_2008_bhaccretion,Martnez:2014}. An accretion disk is a flattened shape composed of revolving gaseous components spiraling into a massive core body. In the presence of interstellar materials, accretion disks form around black holes. Black hole accretion disks form when gaseous materials rotate in unstable constrained orbits \cite{Kato_2008_bhaccretion,Martnez:2014}. Under certain situations, gas particles fall into the gravitational potential of black holes, producing gravitational energy in the form of heat. The inner portion of the accretion disk cools when some heat is converted into radiation \cite{Kato_2008_bhaccretion,Martnez:2014}. When radio, optical, or X-ray telescopes detect the outgoing radiation, the electromagnetic spectrum can be examined. The parameters of radiation are determined by the velocity of gas particles, which can also be influenced by the center mass's structure. In this regard, studying the geodesic structure of particles near black holes, particularly photon orbits ($r_{ph}$) and innermost stable circular orbits ($r_{_{ISCO}}$), is a promising area for further research. These radii play a crucial role in understanding black hole accretion disks. In thin accretion disks, the inner edge corresponds with the innermost stable circular orbit (ISCO). Spacetime has an impact on these radii's positions and the geodesic structure of particles. In Refs. \cite{2024JHEAp..44..279M,2024JHEAp..43...51D,2024JHEAp..43....1S,2024EPJC...84..988R,Ditta_2023_Constraining study,2024PDU....4601689J,Feng:2024zxi} different black holes spacetime properties and their particle geodesic motions and accretions have been investigated extensively. The efficiency of converting rest-mass energy into radiative energy can be estimated using these radii \cite{xie.2012}. Studying accretion disk emission spectra can yield valuable astrophysical data. The accretion disks of black holes have received significant interest and have been researched extensively in literature \cite{Michel:1972,Jamil:2008bc,Babichev:2008dy,JimenezMadrid:2005rk,Babichev:2008jb,Giddings:2008gr,Sharif:2011ih,John:2013bqa,Debnath:2015yva,Ganguly:2014cqa,Mach:2013fsa,
Mach:2013gia,Karkowski:2012vt,Yang:2015sfa,Babichev:2004yx,Babichev:2005py,Gao:2008jv,Barausse:2018vdb,Nozari:2020swx,Zheng:2019mem,Salahshoor:2018plr,Jiao:2016iwp,Uniyal_2024_nonlinearly,Uniyal_2023}.\\

In an accretion disk, particles move in stable orbits. However, when perturbed by restoring forces, they can oscillate vertically and radially with epicyclic frequencies.
The inner portion of an accretion disk is characterized by oscillations that occur in reaction to perturbations. Oscillations can cause significant and chaotic time changes in the spectra of such systems. Studying orbital and epicyclic frequencies (radial and vertical) is crucial for understanding the physics of relativistic accretion disks surrounding black holes. Isper \cite{Ipser_1994,Ipser_1996}, Wagoner \cite{Wagoner_1999}, Kato \cite{Kato_2001}, and Ortega-Rodriguez et al. \cite{Ortega-rodriguez_2008} have all conducted research in this field. Kluzniak and Abromowicz \cite{Kluzniak_2001} postulated a resonance between frequency modes as a physical explanation for existing Quasi-Periodic Oscillations (QPOs). Several studies, including van der Klis \cite{Klis_2000} and McClintock et al. \cite{McClintock_2004}, have examined QPOs in the X-ray fluxes of astrophysical phenomena like neutron stars and black holes. Johannsen and Pradhan \cite{Johannsen_2013,Pradhan_2019} analyzed the radial and vertical epicyclic frequencies in the Kerr-like metric. Also there are many other black holes spacetime \cite{2024JHEAp..44..172C,2024JHEAp..44...99C,Javed_2024_Joule-Thomson expansion,Kundu_2024_Gravitational lensing,Chaudhary_2024_Addressing,Feng:2024tgc,Ditta:2024tdo} in which the authors examine quasi-periodic oscillations and uncovering the dynamical features of these systems.\\

Based on theoretical and observational investigations, the presence of an accretion disk around black holes appears to be the principal source of information regarding gravity and geometry in the strong field regime. It can also enable highly powerful testing to probe undiscovered aspects of high energy phenomena occurring in the close proximity of black holes. As a result, it is worthwhile to consider important and valuable insights into spacetime geometry in the strong field regime, as these can have a significant impact on test particles' geodesics and change observable quantities such as the innermost stable circular orbit (ISCO) and accretion disk parameters. With these in mind, having confidence in the conclusions drawn from observations on accretion disks is critical. Thus, observational data on the thin accretion disk, as well as the expected thermal spectra, could play an important role in proving gravity in the strong field regime. Besides accretion disks can help to distinguish between different modified theories of black holes through several mechanisms. Different modified theories of gravity may predict varying behaviors for test particles' geodesics near the black holes \cite{Mustafa_2023_Circular orbits and accretion disk,Mukherjee:2024hht,Rehman_2023_Matter accretion onto,Mukherjee_2024_The estimation of}. By analyzing the motion of particles in the accretion disk, particularly their orbits, we can identify deviations from predictions made by general relativity. The location of the ISCO is sensitive to the underlying theory of gravity. Modified theories may alter the expected radius of the ISCO \cite{2024JHEAp..44..279M,2024JHEAp..43...51D,2024JHEAp..43....1S,2024EPJC...84..988R,Ditta_2023_Constraining study,2024PDU....4601689J,Feng:2024zxi}. Observations of radiation from the accretion disk can provide precise measurements of this radius, allowing us to test these theories against empirical data. The thermal spectra emitted by accretion disks can vary based on the spacetime geometry predicted by different theories \cite{Mustafa_2023_Circular orbits and accretion disk,Mukherjee:2024hht,Rehman_2023_Matter accretion onto,Mukherjee_2024_The estimation of,Abbas_2023_Accretion disk around RN,Atamurotov:2024nre}. By comparing observed spectra with theoretical models, we can identify inconsistencies that may suggest the need for modifications to general relativity or support alternative theories. With this goal in mind, we investigate the features of a novel solution of a $4$D charged black hole in Einstein-Gauss-Bonnet theory. We study the remarkable characteristics and radiation properties of this black hole's accretion disk. In addition, we investigate the ISCOs around a charged black hole in EGB theory.

The paper is organized as follows: In Section II, we provide an overview of the proposed $4D$ EGB gravity and associated charged static spherically symmetric black hole solution. In Section III, we examine how test particles move in the $4D$ charged EGB black hole spacetime. To simplify our investigation, we focus on the equatorial plane in a polar coordinate system. To investigate circular orbits, the effective potential is obtained using this configuration. We investigate the positions of various characteristic radii in equatorial plan, including marginally stable circular orbits ($r_{_{ISCO}}$), and photon orbits ($r_{ph}$). Then, we derive the radial
and vertical epicyclic frequencies. Also, we intend to investigate the properties of accretion disk around a $4$D charged EGB black hole such as the radiative efficiency, energy flux, temperature profile, and differential luminosity. In Section IV, we consider static spherically symmetric accretion. In Section V, we present our Summary and Conclusions.

\section{Charged black hole solutions in $4D$ EGB thoery}

Lovelock's theorem states that the GB term only affects gravitational dynamics when $D$ is greater than $4$. Glavan and Lin \cite{Glavan_2020} demonstrate that rescaling the GB coupling constant, $\alpha \to \frac{\alpha }{{D - 4}}$, with the limit $D \to 4$ in the GB term, is one way to evaluate this non-trivial contribution at $D=4$. By avoiding Lovelock's theorem, Glavan and Lin \cite{Glavan_2020} were able to construct a $4D$ EGB BH solution.\\

Glavan and Lin's limiting approach is insufficient to account for Gauss-Bonnet gravity's degrees of freedom, requiring extra scalar degrees of freedom in the framework \cite{Li_2020,Bonifacio2020,Fernandes20201,Kobayashi_2020,Hennigar_2020}. A pure gravity theory with Gauss-Bonnet terms remains elusive. Aoki, Gorji, and Mukhohyama \cite{Aoki_2020} solved the difficulty of obtaining a consistent theory of Gauss-Bonnet gravity in $4D$ by dimensional reduction of higher-dimensional Gauss-Bonnet gravity, which is given by the following action,

\begin{equation}\label{2}
S = \frac{1}{{2{\kappa ^2}}}\int {{d^D}x\sqrt { - g} \left[ {R + \alpha {\cal G}} \right]} \,,
\end{equation}
In this equation, ${\cal G}=R_{\mu\nu\lambda\sigma} R^{\mu\nu\lambda\sigma}-4 R_{\mu\nu} R^{\mu\nu}+R^{2}$ represents the Gauss-Bonnet combination of curvature squared terms, $g_{\mu\nu}$ is the $D = d + 1$ dimensional metric, and $\kappa$ is Newton's constant in the $D$ dimension. In \cite{Aoki_2020} the authors demonstrate that the $d \to 3$ limit can be smooth with additional restrictions. In the limit $d \to 3$, the gravitational solutions of $D = d + 1$ theory that follow these restrictions also satisfy $4D$ Gauss-Bonnet gravity.\\
In \cite{Fernandes_2020_2}, a spherically symmetric charged back hole solution of 4D Gauss-Bonnet gravity was derived using the following line element

\begin{equation}\label{4}
d{s^2} =  - F(r)d{t^2} + \frac{1}{{F(r)}}d{r^2} + {r^2}d{\Omega ^2}\,,
\end{equation}
in which

\begin{equation}\label{5}
F\left( r \right) = 1 + \frac{{{r^2}}}{{2\alpha }}\left[ {1 \pm \sqrt {1 + 4\alpha \left( {\frac{{2M}}{{{r^3}}} - \frac{{{Q^2}}}{{{r^4}}}} \right)} } \right]\,,
\end{equation}
$M$ is the mass and $Q$ represents electric charge. \\

Glavan and Lin's static black hole solutions match with the \cite{Aoki_2020}. Additionally, the charged black hole metric (\ref{4}) provides a consistent solution for $4D$ Gauss-Bonnet gravity coupled to  Maxwell's electrodynamics.\\
Eq. (\ref{5}) yields two branches of black hole solutions. However, we will focus on the minus sign of the solution because of its attractive massive source \cite{Dadhich2007,Torii2005}.\\
Now consider the horizon of $4D$ charged EGB BH. The positive root of $F(r) = 0$ yields:
\begin{eqnarray}\label{Eq:hor}
r_{h}=M + \sqrt{M^2-Q^2-\alpha }\, .
\end{eqnarray}

The preceding equation clearly demonstrates that an extremal black hole is given by with horizon $r=M$ if $\alpha+Q^2=M^{2}$ is met. If the term $\alpha+Q^2$ dominates over $M^2$, the black hole becomes a naked singularity. Yang et al. \cite{Yang_2020} addressed this issue of the $4D$ charged EGB black hole. In \cite{Atamurotov2021} authors illustrated how the black hole horizon varies with $\alpha$ and $Q$. They showed that the GB coupling constant and black hole charge have a similar impact in shifting the outer horizon towards a lower $r$, i.e. to the core singularity. This is consistent with the observation that the GB coupling constant has a repulsive gravitational effect \cite{Dadhich20202,Shaymatov2020}.

\section{Motion of a test particle in charged $4D$ EGB black hole spacetime}\label{MoTP}

The geodesic structure of the spacetime is ruled by the path of a test particle \cite{Rezzolla_2014}. Using the Lagrangian formalism, we can scrutinize time-like geodesics around 4D charged EGB black hole \cite{Salahshoor:2018plr}.

The line element of the test particle remains invariant with time reversal and axial symmetries \cite{Nozari:2020swx}. Therefore, space-time in 4D charged EGB gravity is static and symmetric and has two Killing vectors ${\zeta _t} = \frac{\partial }{{\partial {x^\mu }}}\left( {1,0,0,0} \right) = {\partial _t}\,$ and ${\zeta _\varphi } = \frac{\partial }{{\partial {x^\mu }}}\left( {0,0,0,1} \right) = {\partial _\varphi }\,$, \cite{Nozari_2023,Hussain_2015}. By using these two Killing vectors, we can derive the two conserved quantities of energy ($E$) and angular momentum ($L$) for the path of the test particle with the four-velocity ${v^\mu } = \left( {{v^t},{v^r},{v^\theta },{v^\varphi }} \right)\,$, as follows,
\begin{equation}\label{6}
E =  - {g_{\mu \nu }}\zeta _t^\mu {v^\nu } \equiv  - {v_t}\,,
\end{equation}
\begin{equation}\label{7}
L =  - {g_{\mu \nu }}\zeta _\varphi^\mu {v^\nu } \equiv  - {v_\varphi}\,.
\end{equation}

Using the four-velocity-vector normalization condition (${v^\mu }{v_\mu } = 1\,$) and the Euler-Lagrange equation ($\frac{d}{{dt}}\left( {\frac{{\partial {\cal L}}}{{\partial {{\dot x}^\mu }}}} \right) - \frac{{\partial {\cal L}}}{{\partial {x^\mu }}} = 0\,$), we get the following equation
\begin{equation}\label{8}
\left[ {\frac{{{{\left( {{v^r}} \right)}^2}}}{{1 + \frac{{{r^2}}}{{2\alpha }}\left[ {1 - \sqrt {1 + 4\alpha \left( {\frac{{2M}}{{{r^3}}} - \frac{{{Q^2}}}{{{r^4}}}} \right)} } \right]}} + {r^2}{{\left( {{v^r}} \right)}^2}} \right] = \left[ {\left( {\frac{{{r^2}}}{{2\alpha }}\left[ {1 - \sqrt {1 + 4\alpha \left( {\frac{{2M}}{{{r^3}}} - \frac{{{Q^2}}}{{{r^4}}}} \right)} } \right]} \right){{\left( {{v^t}} \right)}^2} - {r^2}{{\left( {{v^\varphi }} \right)}^2}{{\sin }^2}\theta } \right]\,.
\end{equation}

\begin{figure}[htb]
\centering
\subfloat[\label{FigUeffa} for $\alpha= 0.3$]{\includegraphics[width=0.475\textwidth]{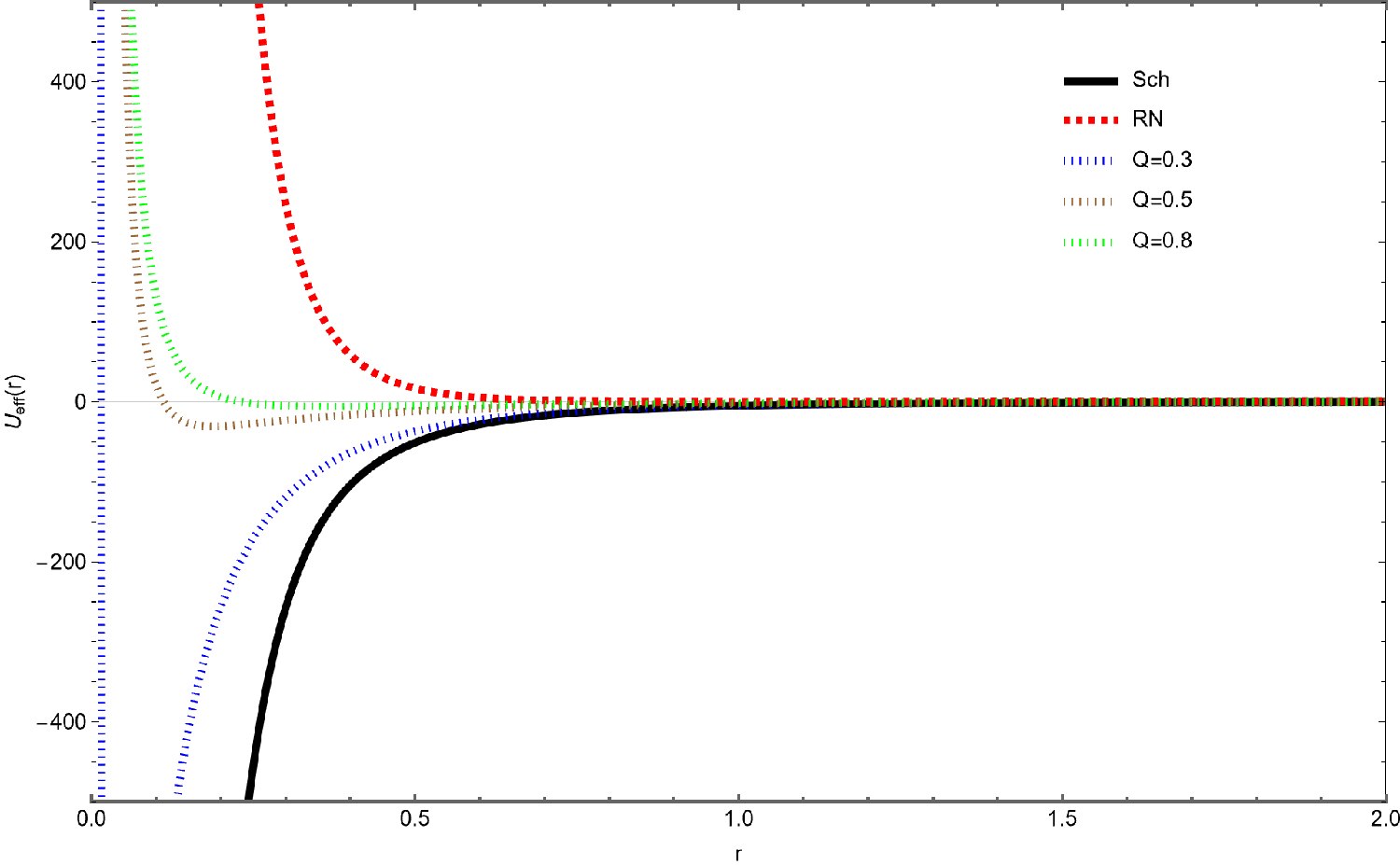}}
\,\,\,
\subfloat[\label{FigUeffb} For $Q =0.3$]{\includegraphics[width=0.475\textwidth]{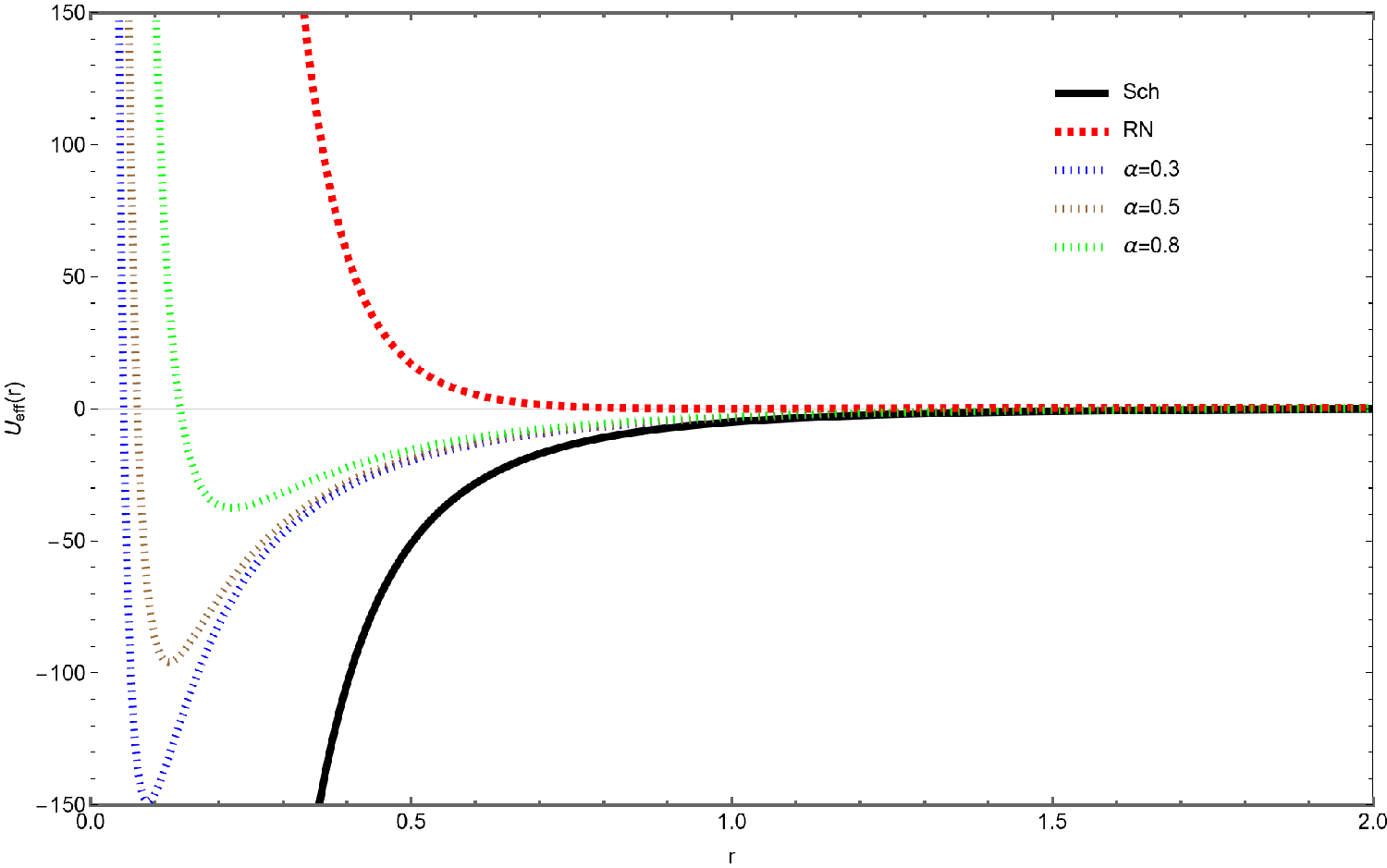}}
\caption{\label{FigUeff}\small{\emph{The plot of effective potential of the charged $4D$ EGB black hole versus the radial coordinate for different values of ${\alpha}$ and $Q$. The black solid and the red dashed lines are the Schwarzschild and Reissner–Nordström black hole solutions in GR.}}}
\end{figure}

Now we assume that the particle is moving in the equatorial plane with $\theta  = \frac{\pi }{2}\,$. According to the movement of the particle in the equatorial plane, the velocity in the $\theta$ direction should be zero. Therefore, the components of the four-velocity vector for the test particle in 4D charged EGB gravity are derived as follows,
\begin{equation}\label{9}
\left\{ {\begin{array}{*{20}{c}}
  {{v^\theta } = 0} \\
\\
  {{v^\varphi } =  - \frac{L}{{{r^2}}}} \\
\\
  {{v^t} =  - \frac{E}{{1 + \frac{{{r^2}}}{{2\alpha }}\left[ {1 - \sqrt {1 + 4\alpha \left( {\frac{{2M}}{{{r^3}}} - \frac{{{Q^2}}}{{{r^4}}}} \right)} } \right]}}} \\
\\
  {{v^r} = \sqrt { - \left( {1 + \frac{{{r^2}}}{{2\alpha }}\left[ {1 - \sqrt {1 + 4\alpha \left( {\frac{{2M}}{{{r^3}}} - \frac{{{Q^2}}}{{{r^4}}}} \right)} } \right]} \right)\left( {1 - \frac{{{E^2}}}{{1 + \frac{{{r^2}}}{{2\alpha }}\left[ {1 - \sqrt {1 + 4\alpha \left( {\frac{{2M}}{{{r^3}}} - \frac{{{Q^2}}}{{{r^4}}}} \right)} } \right]}} + \frac{{{L^2}}}{{{r^2}}}} \right)} }.
\end{array}} \right.\
\end{equation}
 By some simple calculations and using the Euler-Lagrange Eq. \eqref{8}, we derive the effective potential equation as
\begin{equation}\label{10}
\begin{gathered}
  {\left( {{v^r}} \right)^2} + {U_{eff}} = {E^2} \hfill \\
\\
  {U_{eff}} = \left( {1 + \frac{{{r^2}}}{{2\alpha }}\left[ {1 - \sqrt {1 + 4\alpha \left( {\frac{{2M}}{{{r^3}}} - \frac{{{Q^2}}}{{{r^4}}}} \right)} } \right]} \right)\left[ {1 + \frac{{{L^2}}}{{{r^2}}}} \right]. \hfill \\ \,
\end{gathered}
\end{equation}
Effective potential analysis is important for analyzing geodesic structure. For circular orbits, the effective potential's local extremum determines their locations. Figure \ref{FigUeff} compares the behavior of the effective potential $U_{eff}$  for the charged $4D$ EGB black hole to the Schwarzschild and Reissner–Nordström scenario in general relativity. The figure shows that the effective potential has a minimum at the photon sphere radius $r_{ph}$ for each $\alpha$ and $Q$ value. As $r$ approaches $\infty$, the effective potential becomes constant. Fig.\ref{FigUeffa} shows that increasing $Q$ increases the effective potential for a charged $4D$ EGB black hole with a fixed $\alpha$. Fig.\ref{FigUeffb}  illustrates that by fixing $Q$, increasing $\alpha$ results in amplifying the black hole's effective potential.

\subsection{Photon spheres and innermost stable circular orbits}

The primary feature of circular orbits is $\dot{r}=\ddot{r}=0$. Alternatively, $v^{r}=\dot{v}^{r}=0$. As a result, Eqs. \eqref{9} show that for circular orbits, $E^{2}=U_{eff}$ and hence $\frac{dU_{eff}}{dr}=0$ are required. Solving these two equations simultaneously and using Eqs. \eqref{5}, and \eqref{9} results in the following relations for the total (specific) energy $E$, total (specific) angular momentum $L$, and the angular velocity $\Omega_{\varphi}\equiv\frac{d\varphi}{dt}=\frac{v^{\varphi}}{v^{t}}$ for the test particle in charged $4D$ EGB black hole background
\begin{equation}\label{11}
\begin{gathered}
  {E^2} = \frac{2F^{2}(r)}{2F(r)-rF'(r)}=\frac{r^2 \sqrt{1-\frac{4 \alpha  \left(Q^2-2 M r\right)}{r^4}} \left(r^2 \left(\sqrt{1-\frac{4 \alpha  \left(Q^2-2 M r\right)}{r^4}}-1\right)-2 \alpha \right)^2}{2 \alpha ^2 \left(2 r \left(r \sqrt{1-\frac{4 \alpha  \left(Q^2-2 M r\right)}{r^4}}-3 M\right)+4 Q^2\right)}
\\
  {L^2} =\frac{r^{3}F'(r)}{2F(r)-rF'(r)}= \frac{r^2 \left(r^4 \left(\sqrt{1-\frac{4 \alpha  \left(Q^2-2 M r\right)}{r^4}}-1\right)-2 \alpha  M r\right)}{\alpha  \left(2 r \left(r \sqrt{1-\frac{4 \alpha  \left(Q^2-2 M r\right)}{r^4}}-3 M\right)+4 Q^2\right)}
\\
  {\Omega^{2}_{\varphi }} = \frac{1}{2r}F'(r)=\frac{r^3 \left(\sqrt{1-\frac{4 \alpha  \left(Q^2-2 M r\right)}{r^4}}-1\right)-2 \alpha  M}{2 \alpha  r^3 \sqrt{1-\frac{4 \alpha  \left(Q^2-2 M r\right)}{r^4}}}. \hfill \\
\end{gathered} \,,
\end{equation}
Eqs. \eqref{11} show that the existence of photon circular orbits requires the condition $2[F(r)][-rF'(r)]>0$ for the total energy and angular momentum to be real.

\begin{figure}[htb]
\centering
\subfloat[\label{FigE2a} $\alpha$ = 0.3]{\includegraphics[width=0.475\textwidth]{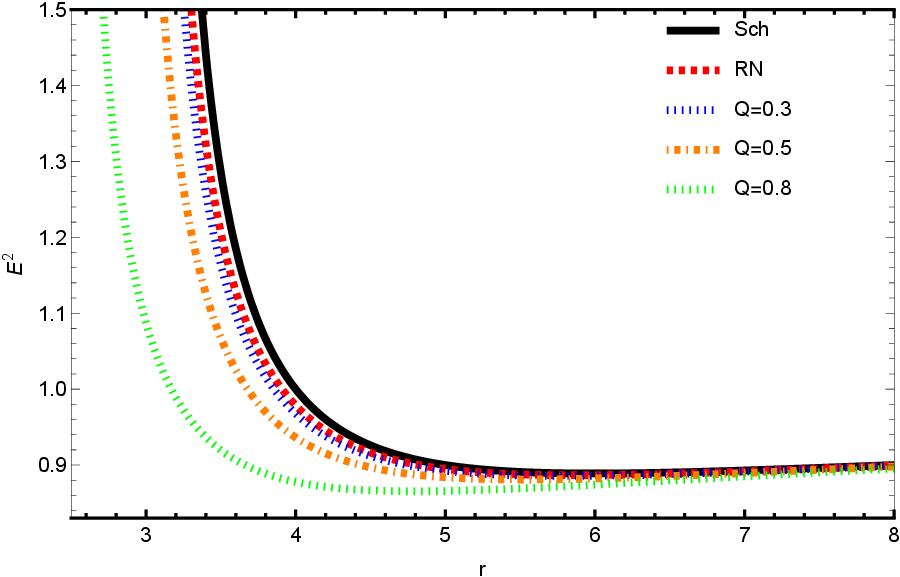}}
\,\,\,
\subfloat[\label{FigE2b} $Q$ = 0.3]{\includegraphics[width=0.475\textwidth]{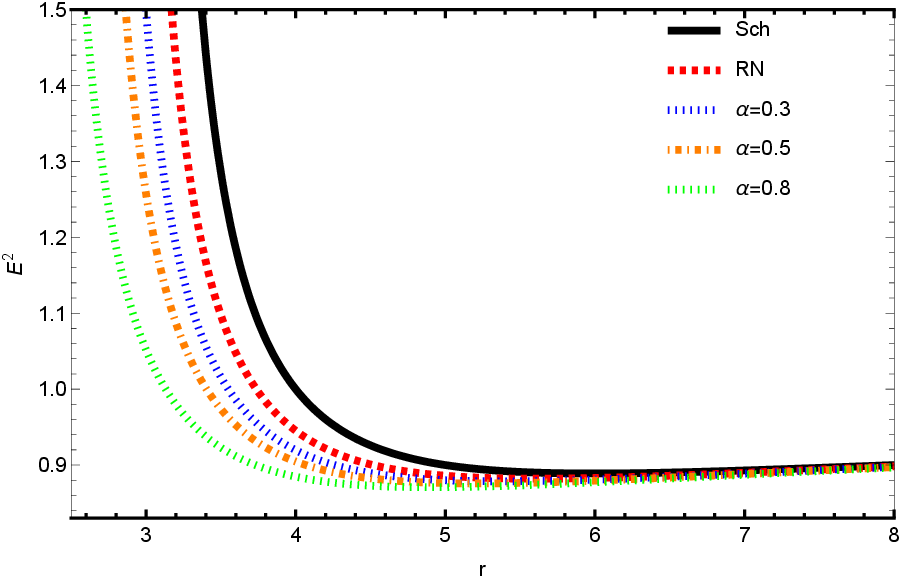}}
\caption{\label{FigE2}\small{The plot of $E^2$ of charged $4D$ EGB black hole versus r for different values of ${\alpha}$ and $Q$. The black solid and the red dashed lines are the Schwarzschild and Reissner–Nordström black hole solutions in GR.}}
\end{figure}

\begin{figure}[htb]
\centering
\subfloat[\label{FigL2a} $\alpha$ = 0.3]{\includegraphics[width=0.475\textwidth]{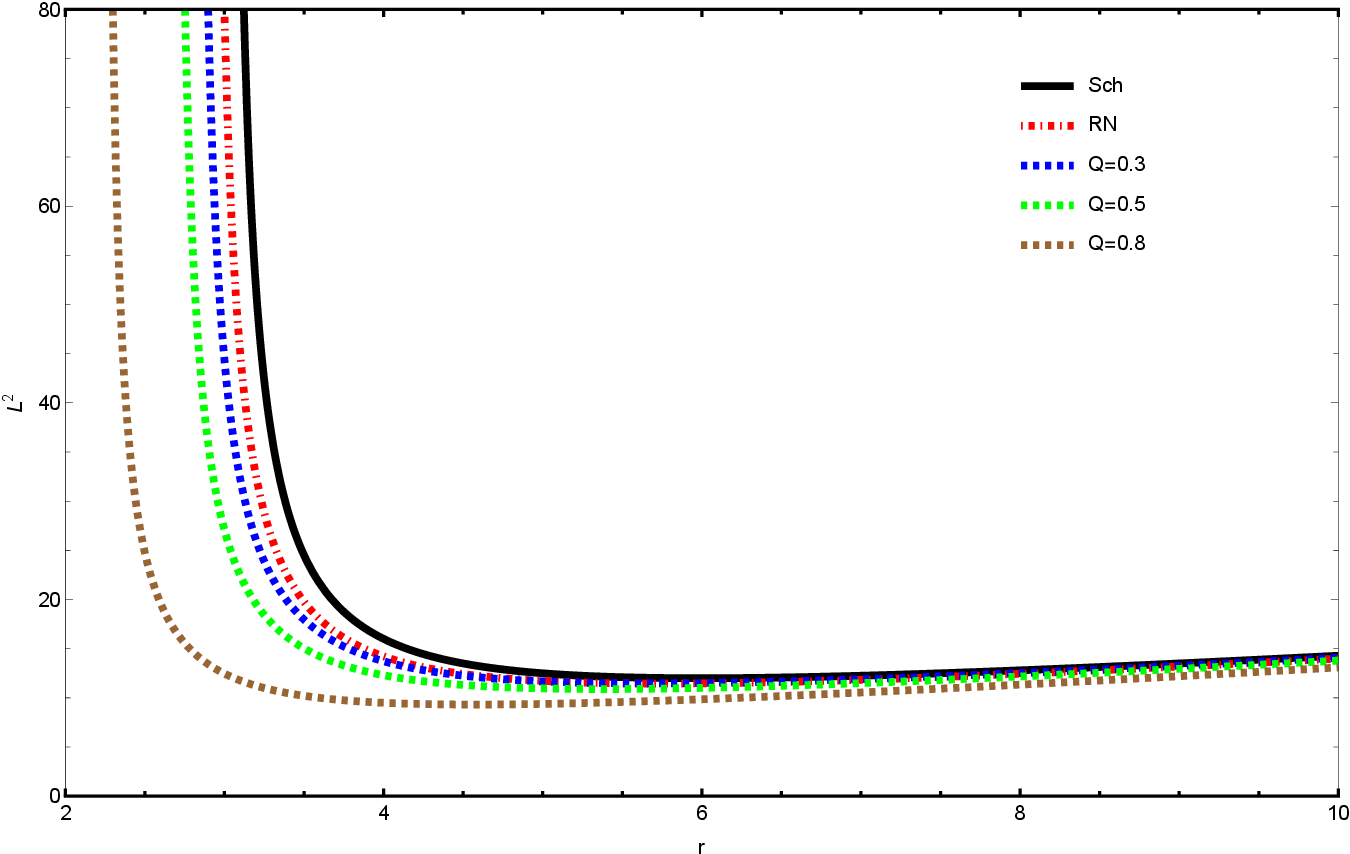}}
\,\,\,
\subfloat[\label{FigL2b} $Q$ = 0.3]{\includegraphics[width=0.475\textwidth]{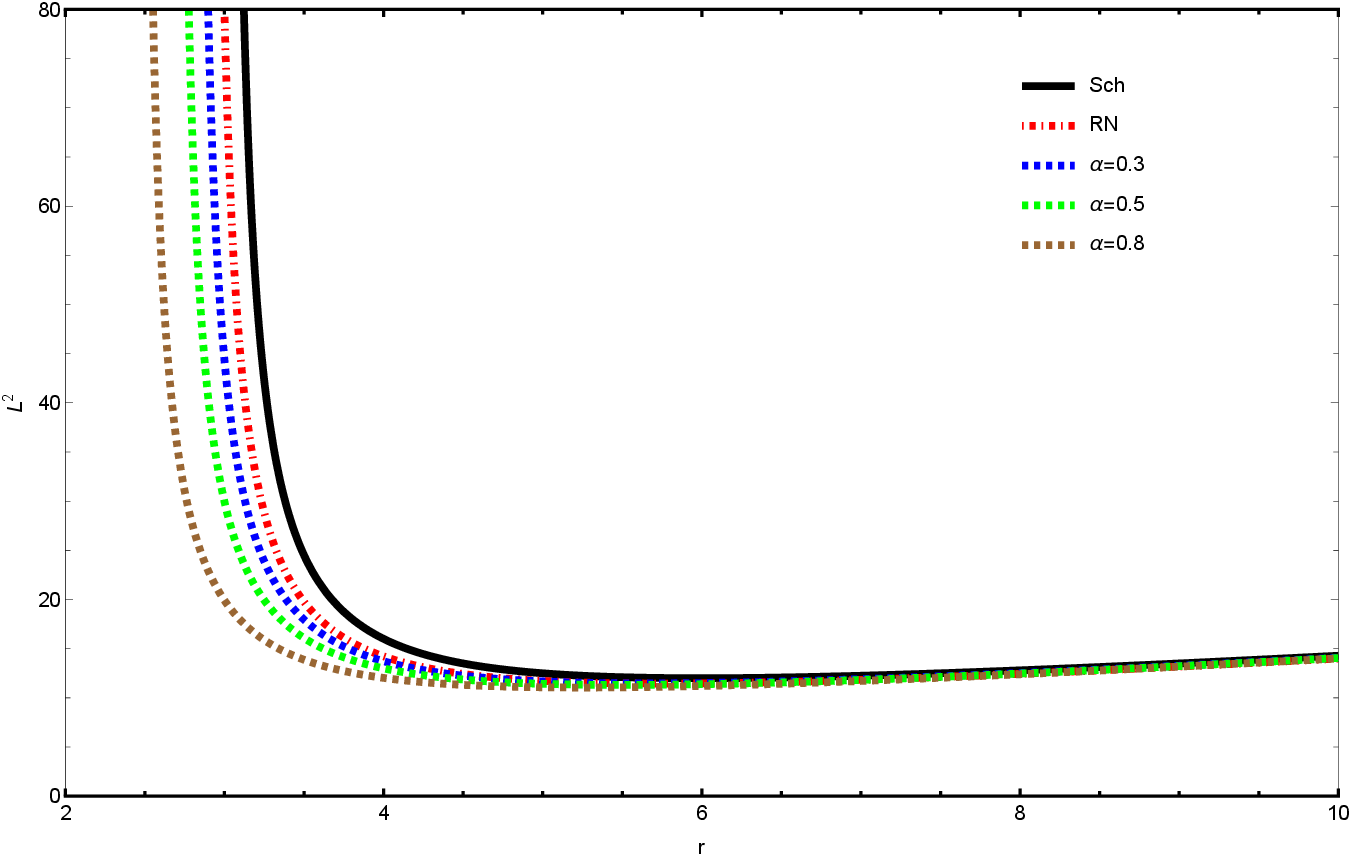}}
\caption{\label{FigL2}\small{The plot of $L^2$ of charged $4D$ EGB black hole versus r for different values of ${\alpha}$ and Q. The black solid and the red dashed lines are the Schwarzschild and Reissner–Nordström black hole solutions in GR}}
\end{figure}

\begin{figure}[htb]
\centering
\subfloat[\label{FigOmega2a} $\alpha$ =0.3]{\includegraphics[width=0.475\textwidth]{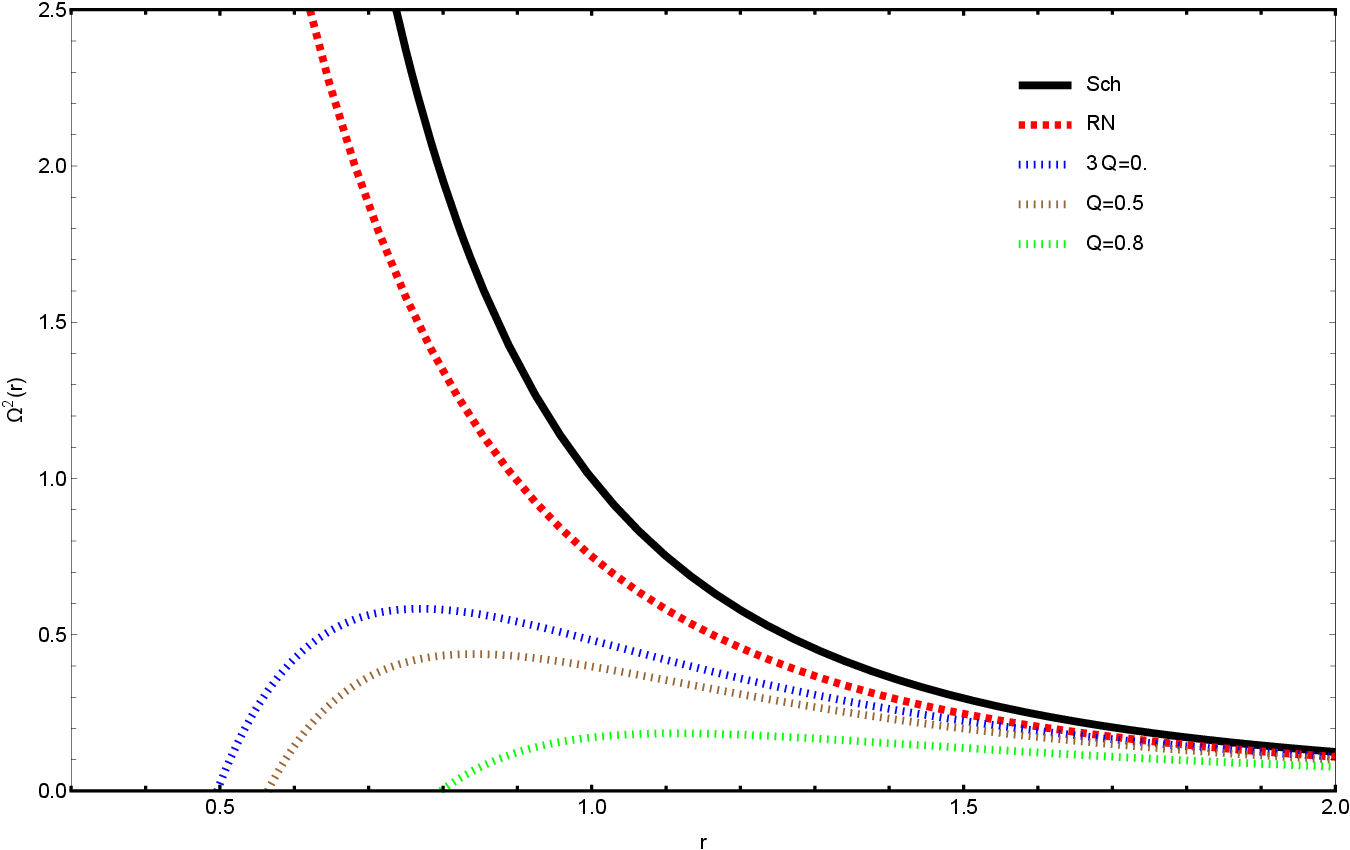}}
\,\,\,
\subfloat[\label{FigOmega2b} $Q$ = 0.3]{\includegraphics[width=0.475\textwidth]{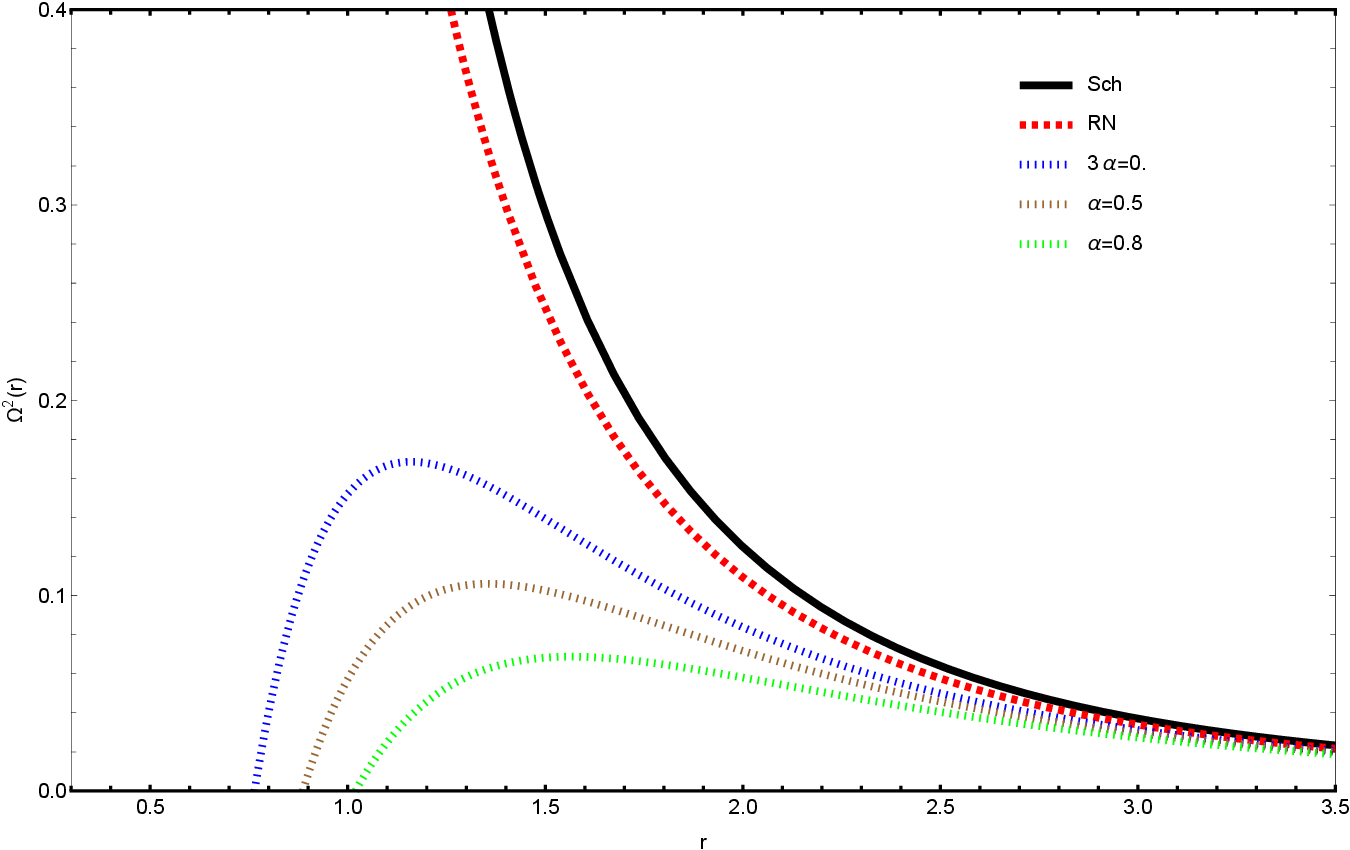}}
\caption{\label{FigOmega2}\small{The plot of $\Omega _\varphi ^2$ of charged $4D$ EGB black hole versus r for different values of ${\alpha}$ and Q. The black solid and the red dashed lines are the Schwarzschild and Reissner–Nordström black hole solutions in GR.}}
\end{figure}

Figure \ref{FigE2} shows the behavior of the $E^{2}$ versus $r$ for different values of $\alpha$ and $Q$, which shows that for a fixed value of $Q$, increasing the parameter $\alpha$ decreases the specific energy of the test particle in the spacetime of the charged $4D$ EGB black hole, whereas the energy becomes almost constant far from the source. Also, by fixing $\alpha$ the specific energy of the test particle reduces by increasing the charge $Q$ in the charged $4D$ EGB black hole spacetime. Figure \ref{FigE2} depicts the Schwarzschild (Sch) and Reissner–Nordström (RN) solutions in GR, which always has much higher values of the quantities than the charged $4D$ EGB black hole situation.
Figure \ref{FigL2} shows the relationship between $L^{2}$ and $r$ for a typical charged $4D$ EGB black hole and the Schwarzschild and Reissner–Nordström solutions in GR for various $\alpha$ and $Q$. Increasing $\alpha$ leads to a decline in $L^{2}$ for fixed values of $Q$, and again, by amplifying the charge, there would be some decreases in $L^{2}$ values when $\alpha$ is constant. These figures show lower values of $E^{2}$ and $L^{2}$ compared to the Schwarzschild and Reissner–Nordström solutions in GR.
The $\Omega_{\varphi}^{2}$ versus $r$ curves for the charged $4D$ EGB black hole are shown in Fig. \ref{FigOmega2} in comparison to the Schwarzschild and Reissner–Nordström cases, where it is observed that for fixed values of $Q$, an increase in $\alpha$ causes the value of $\Omega_{\varphi}^{2}$ to decrease, resulting in curves for the Schwarzschild and Reissner–Nordström cases with higher $\Omega_{\varphi}^{2}$ values than corresponding ones in the charged $4D$ EGB black hole.

The stable circular orbits' locations correspond to the effective potential's local minimum. Thus, an innermost (marginally) stable circular orbit (ISCO) requires the requirements
\begin{equation}\label{1ISCO}
\frac{dU_{eff}}{dr}=0\,,\quad \frac{d^{2}U_{eff}}{dr^{2}}=0\,,
\end{equation}
to be satisfied. The existence of ISCO, $r_{_{ISCO}}$, is entirely a relativistic phenomenon. Unlike classical mechanics, where the effective potential has only one minimum, in GR, the effective potential can yield either a local maximum and minimum or no extremum, depending on the value of $L$ in the effective potential. This extremum is associated with the test particle's stable outer and unstable inner circular orbits. ISCO occurs when the stable and unstable circular orbits overlap for a given value of $L$. The metric coefficient function \eqref{4} is complex, making it impossible to obtain an explicit analytical form of ISCO for the charged $4D$ EGB black hole. Using Wolfram Mathematica (v12.3) to solve equations set \eqref{1ISCO} yields numerical ISCO values for the test particle in the charged $4D$ EGB black hole's spacetime. To accomplish this, we set $M = 1$. Then, for three distinct values of the EGB parameter $\alpha$ and Charge $Q$ in Table \ref{Table1}, we collect the numerical values of $r_{_{ISCO}}$, $L^{2}_{_{ISCO}}$, $E^{2}_{_{ISCO}}$ and $\Omega^2_{_{ISCO}}$ for the charged $4D$ EGB black hole. As it is obvious, the Schwarzschild black hole in GR has an ISCO of $r_{_{ISCO}}=6M$. As seen in Table \ref{Table1}, raising the values of $\alpha$ and $Q$ decreases the ISCO radius for the charged $4D$ EGB black hole. Also, the numerical values of the photon sphere radius are calculated for various values of $\alpha$ and $Q$.

\begin{table}
  \centering
  \begin{tabular}{|p{1.5cm}|p{1.5cm}|p{1.5cm}|p{1.5cm}|p{1.5cm}|p{1.5cm}|p{1.5cm}|p{1.5cm}|}
    \hline
    \multicolumn{7}{|c|}{Circular Motion Parameter} \\
    \hline
    \hline
    $Q$ & ${\alpha}$ & $r_{ISCO}$ & $\eta \%$ & $r_{ph}$ & $E^{2}_{ISCO}$ & $L^{2}_{ISCO}$ & ${\Omega}^{2}_{ISCO}$ \\
    \hline
    \hline
    0 & 0 & 6 & 5.72 & 3 & 0.9428 & 3.4641 & 0.0680 \\
    \hline
    0.3 & 0 & 5.8628 & 11.34 & 2.9388 & 0.8866 & 11.6968 & 0.0049 \\
    \hline
    0.3 & 0.3 & 5.7417 & 11.49 &  2.7874 & 0.8851 & 11.6524 & 0.0052 \\
    \hline
    0.3 & 0.5 & 5.5954 & 11.7 & 2.6696 & 0.883 & 11.4979 & 0.0055 \\
    \hline
    0.3 & 0.8 & 5.3465 & 12.07 & 2.4494 & 0.8793 & 11.2423 & 0.0062 \\
    \hline
    0.5 & 0.3 & 5.497 & 11.93 & 2.6539 & 0.8807 & 11.1368 & 0.0057 \\
    \hline
    0.5 & 0.5 & 5.2101 & 12.42 & 2.5175 & 0.8758 & 10.718 & 0.0065  \\
    \hline
    0.5 & 0.8 &  4.90673 & 12.93 & 2.2383 & 0.8707 & 10.4123 & 0.0076  \\
    \hline
    \end{tabular}
  \caption{The numerical values of $r_{_{ISCO}}$,$\eta \%$,$r_{ph}$, $L^{2}_{_{ISCO}}$, $E^{2}_{_{ISCO}}$ and $\Omega^{2}_{ISCO}$ for a test particle moving in the charged $4D$ EGB black hole spacetime associated with different values of $\alpha$ and $Q$.}\label{Table1}
\end{table}

On the other hand, the Novikov-Thorne model allows for an optically thick and geometrically thin accretion disk near the charged $4D$ EGB black hole. However, because it is expanded horizontally, the accretion disk is extremely thin in its vertical dimension. This results in a tiny vertical size when compared to its vast horizontal extension. Therefore, its height $h$ is substantially smaller than the radius of the disk $r$ extended in the horizontal direction, resulting in $h \ll r$. Because of this attribute of the thin accretion disk created in the surrounding environment of the charged $4D$ EGB  black hole, as well as the thin disk's hydrodynamic equilibrium, vertical entropy and pressure gradients are considered negligible in the accreting matter. The heat generated by stress and dynamical friction cannot be collected in the accretion disk, which is efficiently cooled as seen by thermal radiation on its surface. As a result, the stabilized thin disk is formed, with its inner edge in a stable orbit around the black hole, allowing the plasma to accrete in stable orbits using Keplerian motion. 
We now define the bolometric luminosity of the accretion disk, which is given by
\begin{eqnarray}
\mathcal{L}_{bol}=\eta \dot{M}c^2\, ,
\end{eqnarray}
where $\dot{M}$ and $\eta$ correspond to the rate of matter falling into the black hole and the energy efficiency of the accretion disk, respectively. Astrophysically, it is known that there are various difficulties in observing the bolometric luminosity, which varies based on the black hole properties and physical form. As a result, measuring the bolometric luminosity using theoretical analysis and models becomes increasingly significant. To that purpose, it is useful to define the energy efficiency of the accretion disk around black holes, which is the maximum amount of energy that can be extracted from the accretion disk as compared to matter falling into the black hole. Hence, energy efficiency plays a crucial role in explaining the accretion process, as rest mass-accreting matter emits electromagnetic radiation from the centeral object. The radiation rate of the photon energy emitted from the disk surface can then be used to further calculate the energy efficiency. Therefore, in the event that the photons released from the disk escape to infinity, the efficiency can be estimated using the measured energy at the ISCO, that is
\begin{equation}
    \eta=1-E_{ISCO}\ .
\end{equation}
The radiative efficiency $\eta$ for the photons emitted from the disk can be found using the above calculation for the observed energy at the ISCO. Next, we analyze the energy efficiency of the accretion disk around charged the $4D$ EGB black hole using data from Table \ref{Table1}. Table \ref{Table1} also displays the energy efficiency value. The findings indicate that the radiative efficiency increases due to the influence of EGB parameter $\alpha$ and the charge $Q$.

\subsection{The radiation flux of the accretion disk}\label{RFoAD}

During the accretion process, it is assumed that falling particles at infinity from rest accrete onto the source mass. The gravitational energy released by falling particles is converted into electromagnetic radiation \cite{Kato_2008_bhaccretion,Page:1974he}. The flux of radiant energy over the disk can be described using the following relation in terms of $L$, $E$, and $\Omega_{\varphi}$ \cite{Kato_2008_bhaccretion}
\begin{eqnarray}\label{a13}
K=-\frac{\dot{M}\Omega_{\varphi}'}{4\pi \sqrt{-g}(E-L
\Omega_{\varphi})^{2}}\int_{r_{ISCO}}^{r}(E-L
\Omega_{\varphi})L'dr\,,
\end{eqnarray}
where $\dot{M}$ represents the accretion rate, $\Omega_{\varphi,r}\equiv \frac{d\Omega_{\varphi}}{dr}$, and the parameter $g$ is the determinant of $g_{\mu\nu}$, given by
\begin{eqnarray}\label{a13}
g=det(g_{\mu\nu})=-r^{4}\sin^{2}\theta\,.
\end{eqnarray}
We chose $\sin\theta=1$ to limit our studies to the equatorial plane. From relations \eqref{11}, we would have
\begin{eqnarray}\label{a13}
K(r)=-\frac{\dot{M}}{4\pi r^{4}}\sqrt{\frac{r}{2
F'(r)}}\Big(\frac{[2 F(r)-r F'(r)][r F''(r)-F'(r)]}{[2 F(r)+r
F'(r)]^{2}}\Big)\int_{r_{ISCO}}^{r}\mathcal{F}(r)dr\,,
\end{eqnarray}
where by definition
\begin{eqnarray}\label{a13}
 \mathcal{F}(r)\equiv\sqrt{\frac{r}{2F'(r)}}\frac{[2 F(r)+r
F'(r)][- F''(r)r F(r)+2r F'^{2}(r)-3F'(r) F(r)]}{[2 F(r)-r
F'(r)]^{2}}\,.
\end{eqnarray}

Now, by definition for the charged $4D$ EGB black hole we have

\begin{eqnarray}
  K(r)&=&\frac{\dot{M}\sqrt {\frac{r}{{ - x_{1} + \frac{{r\left( {1 - \sqrt {1 - \frac{{4{Q^2}\alpha }}{{{r^4}}} + \frac{{8M\alpha }}{{{r^3}}}} } \right)}}{\alpha }}}} }{4\sqrt 2 \pi {r^4}{{\left( {r\left( { - x_{1} + \frac{{r\left( {1 - \sqrt {1 - \frac{{4\alpha {Q^2}}}{{{r^4}}} + \frac{{8M\alpha }}{{{r^3}}}} } \right)}}{\alpha }} \right) + 2\left( {1 + \frac{{{r^2}\left( {1 - \sqrt {1 - \frac{{4\alpha {Q^2}}}{{{r^4}}} + \frac{{8M\alpha }}{{{r^3}}}} } \right)}}{{2\alpha }}} \right)} \right)}^2}} \nonumber \\
  && \times \frac{\left( {x_1} - \frac{{r\left( {1 - \sqrt {1 - \frac{{4{Q^2}\alpha }}{{{r^4}}} + \frac{{8M\alpha }}{{{r^3}}}} } \right)}}{\alpha } + r\left( { - {x_1} + \frac{{\left( {{x_1} - \frac{{\left( { - \frac{{80{Q^2}\alpha }}{{{r^6}}} + \frac{{96M\alpha }}{{{r^5}}}} \right){r^2}}}{{2\sqrt {1 - \frac{{4{Q^2}\alpha }}{{{r^4}}} + \frac{{8M\alpha }}{{{r^3}}}} }}} \right)}}{{2\alpha }} + \frac{{1 - \sqrt {1 - \frac{{4{Q^2}\alpha }}{{{r^4}}} + \frac{{8M\alpha }}{{{r^3}}}} }}{\alpha }} \right) \right)}{4\sqrt 2 \pi {r^4}{{\left( {r\left( { - {x_1} + \frac{{r\left( {1 - \sqrt {1 - \frac{{4{Q^2}\alpha }}{{{r^4}}} + \frac{{8M\alpha }}{{{r^3}}}} } \right)}}{\alpha }} \right) + 2\left( {1 + \frac{{{r^2}\left( {1 - \sqrt {1 - \frac{{4{Q^2}\alpha }}{{{r^4}}} + \frac{{8M\alpha }}{{{r^3}}}} } \right)}}{{2\alpha }}} \right)} \right)}^2}} \nonumber \\
  && \times \frac{\left( - r\left( { - {x_1} + \frac{{r\left( {1 - \sqrt {1 - \frac{{4{Q^2}\alpha }}{{{r^4}}} + \frac{{8M\alpha }}{{{r^3}}}} } \right)}}{\alpha }} \right) + 2\left( {1 + \frac{{{r^2}\left( {1 - \sqrt {1 - \frac{{4{Q^2}\alpha }}{{{r^4}}} + \frac{{8M\alpha }}{{{r^3}}}} } \right)}}{{2\alpha }}} \right) \right)}{4\sqrt 2 \pi {r^4}{{\left( {r\left( { - {x_1} + \frac{{r\left( {1 - \sqrt {1 - \frac{{4{Q^2}\alpha }}{{{r^4}}} + \frac{{8M\alpha }}{{{r^3}}}} } \right)}}{\alpha }} \right) + 2\left( {1 + \frac{{{r^2}\left( {1 - \sqrt {1 - \frac{{4{Q^2}\alpha }}{{{r^4}}} + \frac{{8M\alpha }}{{{r^3}}}} } \right)}}{{2\alpha }}} \right)} \right)}^2}}\int_{r_{ISCO}}^{r}\mathcal{F}(r)dr\,,
\end{eqnarray}

where

\begin{equation}
x_{1}=\frac{r^2 \left(\frac{16 \alpha  Q^2}{r^5}-\frac{24 \alpha  M}{r^4}\right)}{4 \alpha  \sqrt{\frac{8 \alpha  M}{r^3}-\frac{4 \alpha  Q^2}{r^4}+1}}
\end{equation}

and

\begin{eqnarray}\label{fr}
  \mathcal{F}\left(r\right)&=&\frac{{\sqrt { - \frac{{r\alpha }}{{ - 2\alpha  + {r^2}\left( { - 1 + x_{1}} \right)}}} \left( {2{Q^2}\alpha  - 5Mr\alpha  + {r^2}\alpha x_{1} + {r^4}\left( { - 1 + x_{1}} \right)} \right)}}{{2{\alpha ^2}\left( {{r^4} - 4\alpha {Q^2} + 8M\alpha r} \right){{\left( {2{Q^2} + r\left( { - 3M + rx_{1}} \right)} \right)}^2}}} \nonumber \\
  && \times \frac{{ {32{Q^4}r\alpha  - 4{r^7}\left( { - 1 + x_{1}} \right) + 4{M^2}r\alpha \left( {10\alpha  - 3{r^2}\left( { - 7 + x_{1}} \right)} \right) + M{r^4}\left( {46\alpha  - 32\alpha x_{1} - 15{r^2}\left( { - 1 + x_{1}} \right)} \right)}}}{{2{\alpha ^2}\left( {{r^4} - 4\alpha {Q^2} + 8M\alpha r} \right){\left( {2{Q^2} + r\left({-3M+rx_{1}}\right)} \right)^2}}} \nonumber \\
  && \times \frac{4{Q^2}\left( { - 6M{\alpha ^2} + 3M{r^2}\alpha \left( { - 9 + x_{1}} \right) + 3{r^5}\left( { - 1 + x_{1}} \right) + 2{r^3}\alpha \left( { - 3 + 2x_{1}} \right)} \right)}{{{2{\alpha ^2}\left( {{r^4} - 4\alpha {Q^2} + 8M\alpha r} \right){{\left( {2{Q^2} + r\left( { - 3M + rx_{1}} \right)} \right)}^2}}}},
\end{eqnarray}

in which we have defined

\begin{equation}
x_{2}\equiv \sqrt{1-\frac{4 \alpha  \left(Q^2-2 M r\right)}{r^4}}.
\end{equation}

Using numerical data from Table \ref{Table1}, one can construct an approximate equation for radiation flux. It should also be highlighted that thermodynamic equilibrium is a fundamental need for the model representing the steady-state accretion disk flow. As a result, the radiation released by the accretion disk surface is comparable to the black body spectrum \cite{Perez:2017spz, Kato_2008_bhaccretion,Page:1974he}. This indicates that the energy flux and the effective temperature of the accretion disk can be connected to the well-known Stefan-Boltzman formula $\mathcal{K}=\sigma_{_{SB}}T^{4}$, where $\sigma_{_{SB}}$ is the Stefan-Boltzman constant. As a result, the effective temperature $T$ of the accretion disk can be determined using this law.

\begin{figure}[htb]
\centering
{\includegraphics[width=0.475\textwidth]{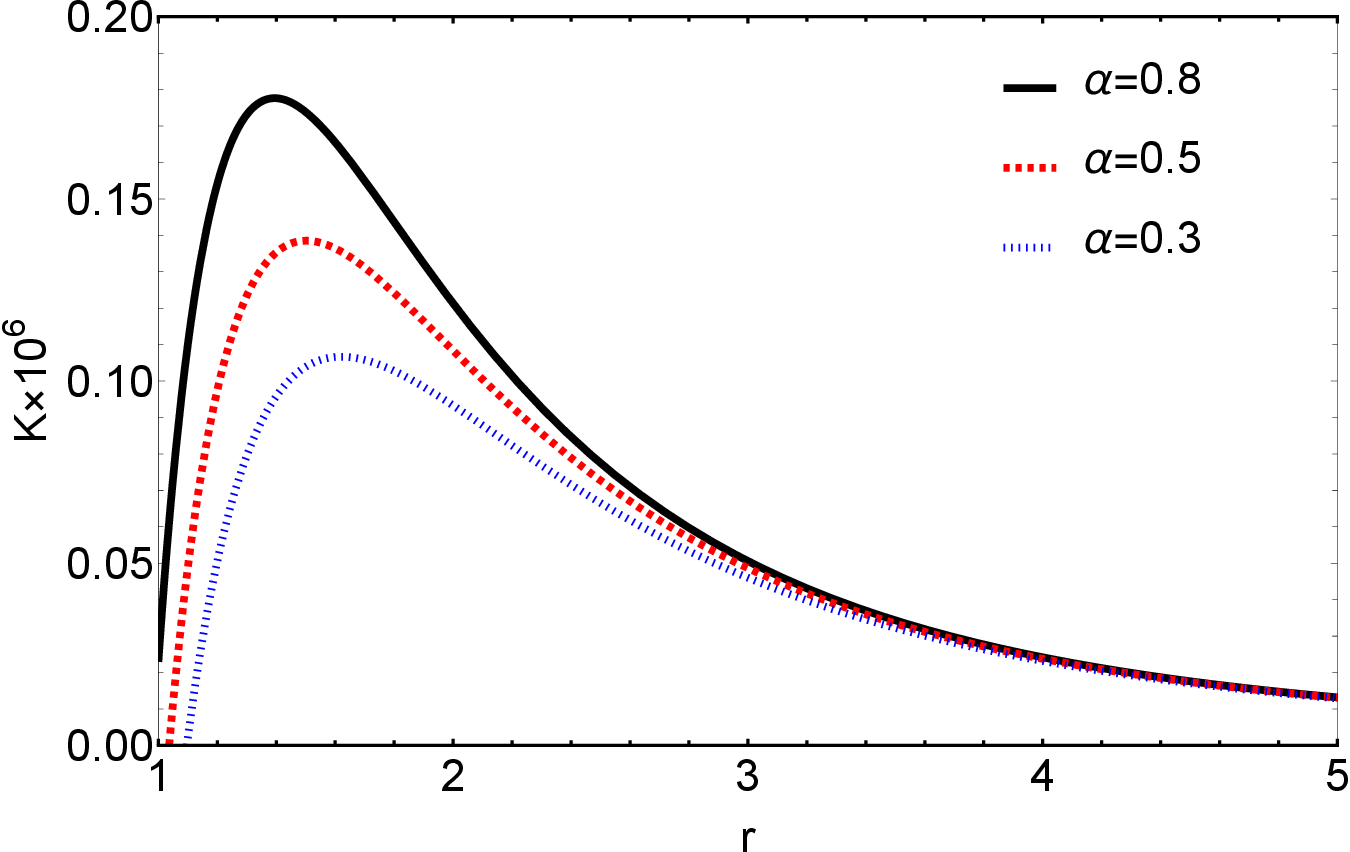}}
\,\,\,
{\includegraphics[width=0.475\textwidth]{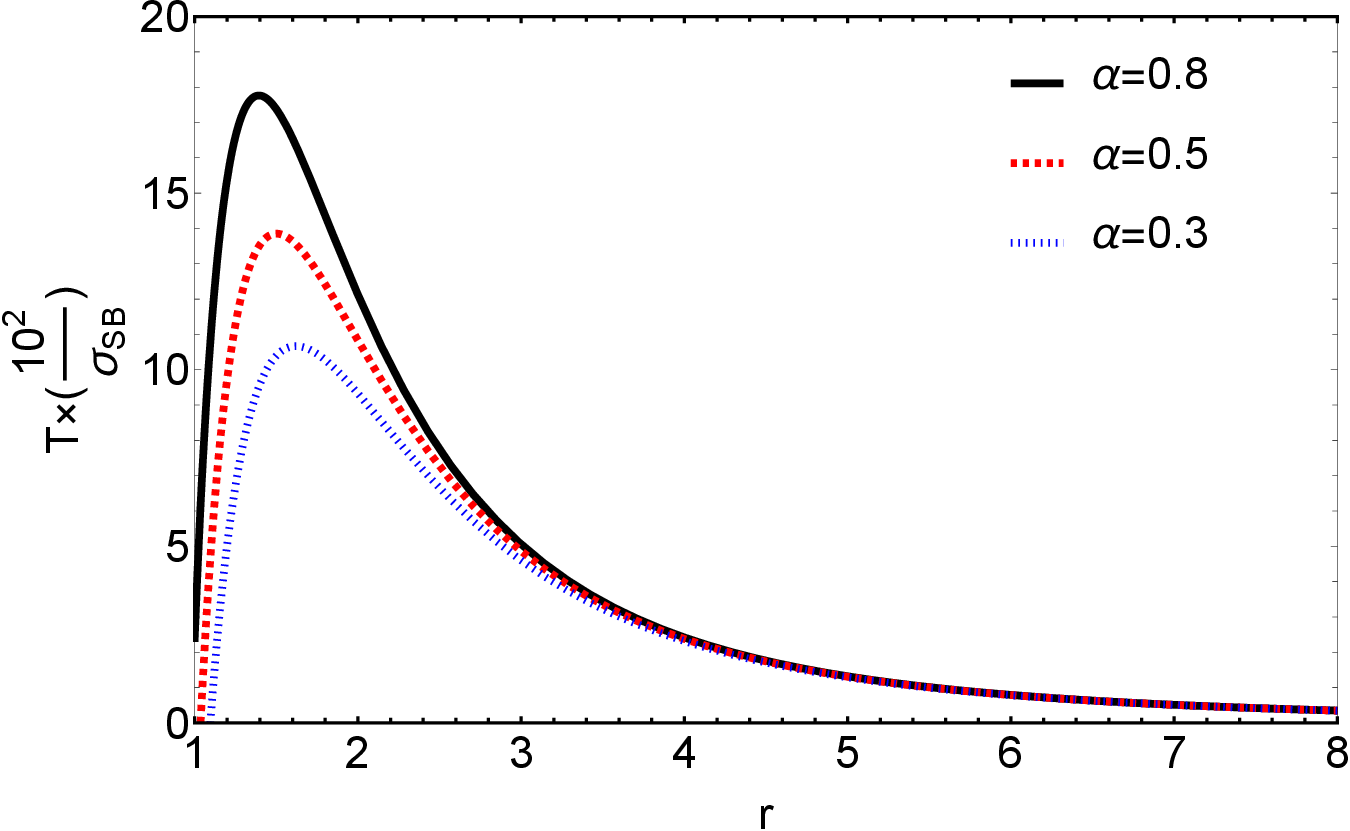}}
\caption{\label{FigT}\small{\emph{The plots of emission rate and temperature $T$ of the charged $4D$ EGB black hole versus radius $r$ for different vales of EGB parameter $\alpha$ ($\alpha = 0.8, 0.5, 0.3$ from top to bottom) for a fixed $Q$.}}}
\end{figure}

Figure \ref{FigT} illustrates the relationship between radiation flux and temperature of the accretion disk for the charged $4D$ EGB black hole in terms of different values of $\alpha$. The radiation flux peaks near the black hole and declines at lower radii. Increasing the EGB parameter $\alpha$ leads to increase energy flux and maximum emission flux at lower radii, whereas the opposite occurs beyond this point. The influence on this parameter is strong near the black hole but weaker farther out. These characteristics are consistent with temperature.\\

\begin{figure}[htb]
\centering
{\includegraphics[width=0.475\textwidth]{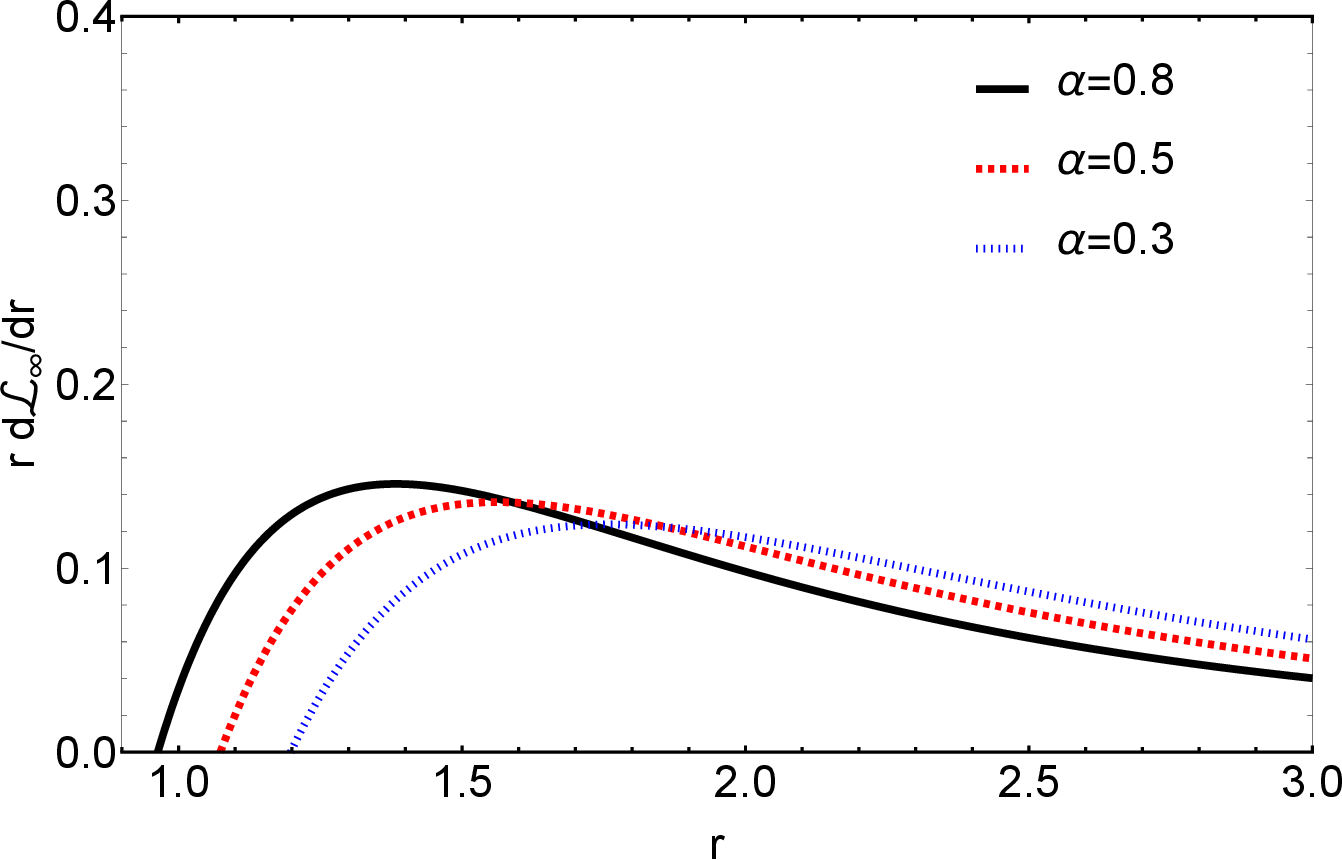}}
\caption{\label{Figlu}\small{\emph{The radial profile of the differential luminosity of the accretion disk of the charged $4$D EGB black hole, on the order of $10^{-2}$ for a fixed $Q$.}}}
\end{figure}

We consider the differential luminosity as one of the key characteristics of the accretion disk, which can contribute to a better understanding of the disk radiation, in addition to the temperature and flux of the accretion disk. The differential luminosity is defined by the following equation,
\begin{equation}\label{Eq:luminosity}
    \dfrac{d \mathcal{L}_{\infty}}{d \ln{r}}=4 \pi r \sqrt{g} E \mathcal{F}(r)\, ,
\end{equation}
in which, $\mathcal{F}(r)$ is given by relation \eqref{fr}.\\
For simplicity, we assume black body radiation can be used to characterize radiation emissions. To do this, we define the spectral luminosity ($\mathcal{L}_{\nu,\infty}$)  in terms of the radiation frequency  ($\nu$) at infinity as
\begin{equation}\label{luminosity2}
    \nu \mathcal{L}_{\nu,\infty}=\dfrac{60}{\pi^3} \int_{r_{ISCO}}^{\infty} \dfrac{\sqrt{g} E}{M_T^2}\dfrac{(u^t y)^4}{\exp\Big[{\dfrac{u^t y}{(M_T^2 \mathcal{F})^{1/4}}}\Big]-1}\, .
\end{equation}
In this equation, $y=h \nu /k T_{\star}$, $k$ and $h$ represent the Boltzmann and Planck constants, respectively, while $M_{T}$ represents total mass. The time component of the four velocity, is denoted by $u^{t}$ and $T_{\star}$ represents the typical temperature associated with the Stefan-Boltzmann law, which is given by
$$\sigma T_{\star}= \dfrac{\dot{M}_0}{4 \pi M_T^2},$$
where $\sigma$ denotes the Stefan-Boltzmann constant. We investigate the differential luminosity of the accretion disk and show its radial profile in Figure \ref{Figlu}. Similar to the accretion disk flux, increasing the EGB parameter $\alpha$ raises differential luminosity. Interestingly, the differential luminosity decreases as distance from the black hole increases. This suggests that the parameter $\alpha$ has a substantial impact solely in the neighborhood of the black hole.

\subsection{Oscillations in accretion disk}\label{OiAD}

Due to the transfer of momentum and energy as well as the presence of dissipation forces, accretion disks need to oscillate in their orbits to reach the equilibrium point. The result of reaching the equilibrium point is horizontal and vertical oscillations in the circular orbits. When the momentum and energy of a particle in an orbit are shifted in the radial direction, the centrifugal force creates an equilibrium state. The inward or outward particle's motion is due to the domination of energy-momentum transfer or centrifugal force, respectively. On the other hand, due to the gravitational field moving in the vertical direction and the form of sinusoidal functions on the particle's movement, we observe oscillation on the geodesic of the particle's in the vertical direction. The sinusoidal nature of this oscillation is an effective factor for reaching the equilibrium point. We represent the frequency of oscillations in the radial and vertical directions with ${\Omega _r }\,$ and ${\Omega _\theta }\,$, respectively. As we mentioned before, resonance between frequencies can cause quasi-periodic oscillations, resulting in chaotic variability in X-ray emissions from numerous galactic black holes. So studying this field is vital in some ways. This section focuses on radial and vertical motions around a circular equatorial plane.\\

The equations for radial and vertical motions are $\frac{1}{2}(\frac{dr}{dt})^{2}=U^{(r)}_{eff}$ and $\frac{1}{2}(\frac{d\theta}{dt})^{2}=U^{(\theta)}_{eff}$. As $v^{\theta}=0$ for radial motion and $v^{r}=0$ for vertical motion, using $v^{r}=\frac{dr}{d\tau}=\frac{dr}{dt}v^{t}$ and $v^{\theta}=\frac{d\theta}{d\tau}=\frac{d\theta}{dt}v^{t}$, we can find
\begin{eqnarray}\label{a13}
&\frac{1}{2}(\frac{dr}{dt})^{2}=-\frac{1}{2}\frac{F(r)}{E^{2}}[1-\frac{E^{2}}{h(r)}+\frac{L^{2}}{r^{2}sin^{2}\theta}]=U^{(r)}_{eff},\nonumber\\
&\frac{1}{2}(\frac{d\theta}{dt})^{2}=-\frac{1}{2}\frac{F^{2}(r)}{r^{2}E^{2}}[1-\frac{E^{2}}{h(r)}+\frac{L^{2}}{r^{2}sin^{2}\theta}]=V^{(\theta)}_{eff}\,.
\end{eqnarray}
To study the radial and vertical epicyclic frequencies, tiny perturbations $\delta r$ and $\delta \theta$ around the circular orbit in the equatorial plane are applied. Using the time derivative of the first equation in \eqref{a13}, the equation of radial oscillations can be described as follows:
\begin{eqnarray}\label{a14}
\frac{d^{2}r}{dt^{2}}=\frac{dU_{eff}^{(r)}}{dr}.
\end{eqnarray}
The perturbed equation of motion for a particle with a deviation $\delta r=r-r_{0}$ from its initial radius at $r=r_{0}$ is as follows:
\begin{eqnarray}\label{a14}
\frac{d^{2}(\delta
r)}{dt^{2}}=\frac{d^{2}U_{eff}^{(r)}}{dr^{2}}(\delta r)\Rightarrow
\ddot{(\delta r)}+\Omega^{2}_{r}(\delta r)=0\,,
\end{eqnarray}
A dot represents a differential with respect to time coordinate $t$, and $\Omega_{r}^{2}\equiv-\frac{d^{2}U_{eff}^{(r)}}{dr^{2}}$. Using the same technique, we can determine the vertical perturbation caused by a deviation $\delta \theta=\theta-\theta_{0}$.
\begin{eqnarray}\label{a14}
\frac{d^{2}(\delta
\theta)}{dt^{2}}=\frac{d^{2}U_{eff}^{(\theta)}}{dr^{2}}\delta
\theta\Rightarrow \ddot{(\delta \theta)}+\Omega^{2}_{\theta}(\delta
\theta)=0\,,
\end{eqnarray}
in which $\Omega^{2}_{\theta}=-\frac{d^{2}}{d\theta^{2}}U^{(\theta)}_{eff}$.
Then, from equations \eqref{a13} in the equatorial plane for the charged $4D$ EGB black hole we would have,

\begin{eqnarray}
  \Omega _r^2&=&\left( {\frac{1}{{2{r^2}{\alpha ^2}{{\left( x_{2} \right)}^4}}}\left( { - 32{M^2}{Q^2}{\alpha ^3} + 16M\left( {4{M^2} + 3{Q^4}} \right)r{\alpha ^3} + 16{Q^2}{r^2}{\alpha ^2}\left( { - 7{M^2}\alpha  + {Q^2}\left( {2 + 4\alpha } \right)} \right)} \right)} \right) \nonumber \\
  && + \left( {\frac{1}{{2{r^2}{\alpha ^2}{{\left( x_{2} \right)}^4}}}\left( { - 4{Q^2}{r^8}\alpha \left( { - 3 + x_{2}} \right) + 4{r^{12}}\left( { - 1 + x_{2}} \right) - 4{r^{10}}\left( {1 + \alpha } \right)\left( { - 1 + x_{2}} \right)} \right)} \right) \nonumber \\
  && + \left( {\frac{1}{{2{r^2}{\alpha ^2}{{\left( x_{2} \right)}^4}}}\left( {4M{Q^2}{r^5}{\alpha ^2}\left( { - 2 + 19x_{2}} \right) + M{r^9}\alpha \left( { - 41 + 25x_{2}} \right)} \right)} \right) \nonumber \\
  && + \left( {\frac{1}{{2{r^2}{\alpha ^2}{{\left( x_{2} \right)}^4}}}\left( {2M{r^7}\alpha \left( {28 - 20x_{2} + \alpha \left( {32 - 23x_{2}} \right)} \right) + 4{r^6}\alpha ( - {M^2}\alpha \left( {17 + 5x_{2}} \right)} \right)} \right) \nonumber \\
  && + \left( {\frac{1}{{2{r^2}{\alpha ^2}{{\left( x{2} \right)}^4}}}\left( {8M{r^3}{\alpha ^2}\left( {4{M^2}\alpha  + {Q^2}\left( { - 20 - 32\alpha  + 4x_{2} + 3\alpha x_{2}} \right)} \right)} \right)} \right) \nonumber \\
  && + \left( {\frac{1}{{2{r^2}{\alpha ^2}{{\left( x_{2} \right)}^4}}}\left( { - 8{r^4}{\alpha ^2}\left( {2{Q^4}\left( { - 1 + 2x_{2}} \right) + {M^2}\left( { - 25 - 32\alpha  + 8x_{2} + 5\alpha x_{2}} \right)} \right)} \right)} \right) \nonumber \\
  && + \left( {\frac{1}{{2{r^2}{\alpha ^2}{{\left( x_{2} \right)}^4}}}\left( {4{r^6}\alpha \left( { - {M^2}\alpha \left( {17 + 5x_{2}} \right) + {Q^2}\left( { - 6 - 8\alpha  + 4x_{2} + 6\alpha x_{2}} \right)} \right)} \right)} \right)
\end{eqnarray}

and
\begin{equation}
\Omega^{2}_{\theta}=\frac{2 \alpha  \left((x_{2}-1) r^4-2 \alpha  M r\right) \left(\frac{r^2 \left(1-\sqrt{\frac{8 \alpha  M}{r^3}-\frac{4 \alpha  Q^2}{r^4}+1}\right)}{2 \alpha }+1\right)^2}{x_{2} r^4 \left((x_{2}-1) r^2-2 \alpha \right)^2}.
\end{equation}

\begin{figure}[htb]
\centering
{\includegraphics[width=0.475\textwidth]{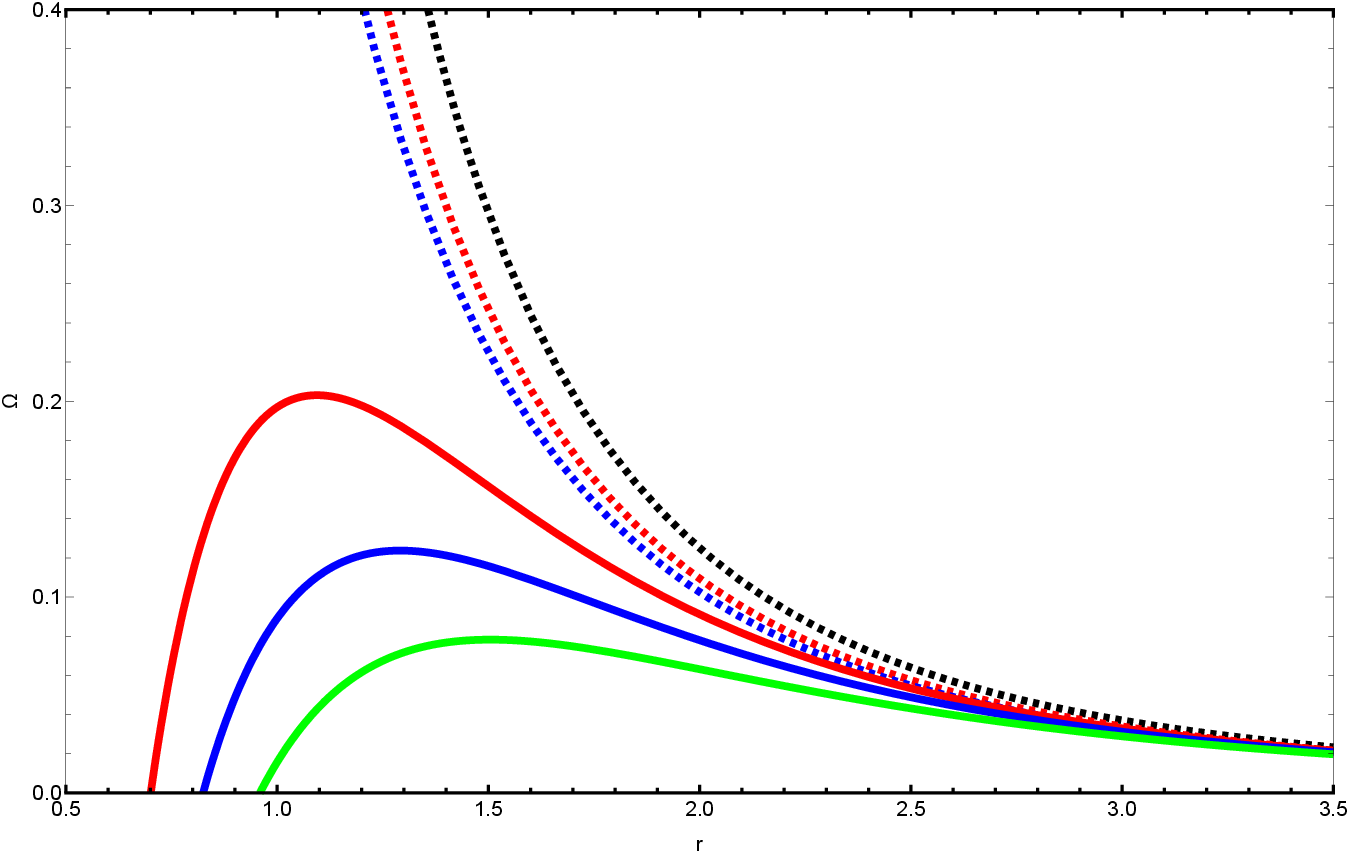}}
\caption{\label{Figfreq}\small{\emph{The plot of epicyclic frequencies of the charged $4D$ EGB black hole for a fixed $Q$. The dashed curves denote vertical frequencies for different values of $\alpha$ ($\alpha = 0.3, 0.5, 0.8$ from top to bottom). The solid curves represent the radial epicyclic frequencies for various values of the EGB parameter $\alpha$ ($\alpha = 0.8, 0.5, 0.3$ from top to bottom).}}}
\end{figure}

Figure \ref{Figfreq} compares epicyclic frequencies vs radius for various values of the EGB parameter $\alpha$. The dashed lines represent the angular and vertical epicyclic frequencies. As $r$ increases, these frequencies decrease. Vertical epicyclic frequencies have a minor decline by increasing the EGB parameter $\alpha$. The radial epicyclic frequency of the charged $4D$ EGB black hole is represented by solid curves. As the parameter $\alpha$ increases, the maximum shifts to smaller radii. Increasing the parameter $\alpha$ leads to higher radial oscillation frequency.
The dependence on this parameter is substantial near the black hole, but weaker away from the center mass. We observe that $\Omega_{r}<\Omega_{\theta}$.

\section{Accretion of Particles onto the charged $4D$ EGB black hole}\label{AoPobh}

In this part, we attempt to find the essential dynamical equations and parameters related with the accretion onto the charged $4D$ EGB black hole using the approach described in Refs. \cite{Nozari:2020swx, Salahshoor:2018plr}. To demonstrate this, we consider spherically symmetric accretion in the equatorial plane with $\theta=\frac{\pi}{2}$. Furthermore, we suppose that the accreting matter is the flowing perfect fluid into the charged $4D$ EGB black hole. It is necessary to mention that a perfect fluid, or ideal fluid, has no viscosity. A perfect fluid lacks internal resistance to flow, allowing it to flow without energy loss from friction among its layers. The notion of a perfect fluid is advantageous in theoretical physics and fluid dynamics for the simplification of equations and models. Thus, in a perfect fluid with negligible viscosity, the transport of angular momentum is primarily influenced by pressure gradients, gravitational interactions, centrifugal forces, and magnetic fields.

\subsection{Dynamical equations}

The equation for the energy-momentum tensor of a perfect fluid is,
\begin{equation}
{T^{\mu \nu }} = \left( {p + \rho } \right){v^\mu }{v^\nu } - p{g^{\mu \nu }}\,.
\end{equation}

Here, due to the symmetries of the problem, the components of the four velocity vectors are in the form of ${v^\mu } = \left( {{v^t},{v^r},0,0} \right){\mkern 1mu} \,$. For simplicity, we assume the geodesics of the particles in the perfect fluid and the perfect fluid geodesics are equivalent and time-like. Therefore, by using the normalization condition ${v^\mu }{v_\mu } = 1\,$, we can reach the following relationship \cite{Salahshoor:2018plr,Nozari:2020swx,Nozari_2023}.
\begin{equation}\label{17}
{v^t} = \frac{\sqrt{F(r)+(v^{r})^{2}}}{F(r)}= \sqrt {\frac{{4\alpha \left( {2\alpha  + {r^2}\left[ {1 - \sqrt {1 - \frac{{4\alpha \left( {{Q^2} - 2Mr} \right)}}{{{r^4}}}} } \right] + 2\alpha {{\left( {{v^r}} \right)}^2}} \right)}}{{{{\left( {2\alpha  + {r^2}\left[ {1 - \sqrt {1 - \frac{{4\alpha \left( {{Q^2} - 2Mr} \right)}}{{{r^4}}}} } \right]} \right)}^2}}}} \,.
\end{equation}

It should be noted that our favorite case is the flowing forward in time; Therefore, we must consider the condition of ${v^t} > 0\,$ while ${v^r} < 0\,$. The equation of the stress-energy tensor conservation is,
\begin{equation}\label{18}
T_{;\mu }^{\mu \nu } = 0 \Rightarrow T_{;\mu }^{\mu \nu } = \frac{1}{{\sqrt { - g} }}{\left( {\sqrt { - g} {T^{\mu \nu }}} \right)_{;\mu }} + \Gamma _{\alpha \mu }^\nu {T^{\alpha \mu }} = 0\,,
\end{equation}

\begin{equation}
\frac{1}{{\sqrt { - g} }}{\left( {\sqrt { - g} {T^{t\nu }}} \right)_{;t}} + \Gamma _{\alpha t}^\nu {T^{\alpha t}} + \frac{1}{{\sqrt { - g} }}{\left( {\sqrt { - g} {T^{r\nu }}} \right)_{;r}} + \Gamma _{\alpha r}^\nu {T^{\alpha r}} = 0\,.
\end{equation}

By using the affine connection of the charged $4D$ EGB black hole in the above equation, we find
\begin{equation}\label{19}
\left( {p + \rho } \right){v^r}{r^2}\sqrt {{{\left( {{v^r}} \right)}^2} + 1 + \frac{{{r^2}}}{{2\alpha }}\left[ {1 - \sqrt {1 + 4\alpha \left( {\frac{{2M}}{{{r^3}}} - \frac{{{Q^2}}}{{{r^4}}}} \right)} } \right]}  = {A_0}\,,
\end{equation}
where ${A_0}$ is an integration constant. On the other hand, we can write the conservation of the energy-momentum tensor for the perfect fluid accreting onto the black hole as
\begin{equation}
{\left( {p + \rho } \right)_{,\nu }}{v_\mu }{v^\mu }{v^\nu } + \left( {p + \rho } \right)v_{;\nu }^\mu {v_\mu }{v^\nu } + \left( {p + \rho } \right){v_\mu }{v^\mu }v_{;\nu }^\nu  + {p_{,\nu }}{g^{\mu \nu }}{v_\mu } + p{v_\mu }g_{;\nu }^{\mu \nu } = 0\,,
\end{equation}
which results in
\begin{equation}\label{20}
\frac{{{\rho ^\prime }}}{{\left( {p + \rho } \right)}} + \frac{{{v^\prime }}}{v} + \frac{2}{r} = 0\,.
\end{equation}
As usuall prime indicates $\frac{\partial }{{\partial {r}}}\,$. By integrating the Eq \eqref{20}, we find

\begin{equation}\label{21}
{r^2}{v^r}{e^{\int {\frac{{d\rho }}{{p + \rho }}} }} =  - {A_1}\,
\end{equation}
where ${A_1}$ is an integraion constant.
As we mentioned above, our favorite case is ${v^r} < 0\,$; therefore, by replacing this condition into the Eq \eqref{20}, we obtain
\begin{equation}\label{22}
\left( {p + \rho } \right)\sqrt {{{\left( {{v^r}} \right)}^2} + 1 + \frac{{{r^2}}}{{2\alpha }}\left[ {1 - \sqrt {1 + 4\alpha \left( {\frac{{2M}}{{{r^3}}} - \frac{{{Q^2}}}{{{r^4}}}} \right)} } \right]} {e^{ - \int {\frac{{d\rho }}{{p + \rho }}} }} = {A_2}\,,
\end{equation}
where ${A_2}$ is also an integration constant.

Now, the mass flux equation is formulated as ${\left( {\rho {v^\mu }} \right)_;} \equiv \frac{1}{{\sqrt { - g} }}{\left( {\sqrt { - g} \rho {v^\mu }} \right)_{,\mu }} = 0\,$. In this case,
\begin{equation}\label{23}
\frac{1}{{\sqrt { - g} }}{\left( {\sqrt { - g} \rho {v^t}} \right)_{,t}} + \frac{1}{{\sqrt { - g} }}{\left( {\sqrt { - g} \rho {v^r}} \right)_{,r}} = 0\,.
\end{equation}

With the assumption being on the equatorial plane ($\theta  = \frac{\pi }{2}\,$), therefore we have
\begin{equation}\label{24}
\rho {v^r}{r^2} = {A_3}\,,
\end{equation}
where ${A_3}$ is another integration constant.

\subsection{Dynamical parameters}

By choosing an isothermal fluid, we can write the equation of state as $p = \omega \rho \,$, where $\omega$ is the equation of state parameter \cite{Salahshoor:2018plr,Nozari:2020swx,Nozari_2023}. In this case, by integrating Eq \eqref{19},\eqref{21},\eqref{22} and \eqref{24}, we obtain
\begin{equation}\label{25}
(\frac{{p + \rho }}{\rho })\sqrt {{{\left( {{v^r}} \right)}^2} + 1 + \frac{{{r^2}}}{{2\alpha }}\left[ {1 - \sqrt {1 + 4\alpha \left( {\frac{{2M}}{{{r^3}}} - \frac{{{Q^2}}}{{{r^4}}}} \right)} } \right]}  = {A_4}\,,
\end{equation}
where ${A_4}$ is an integration constant.

\begin{figure}[htb]
\centering
\subfloat[\label{FigVra} $\alpha$ = 0.3]{\includegraphics[width=0.475\textwidth]{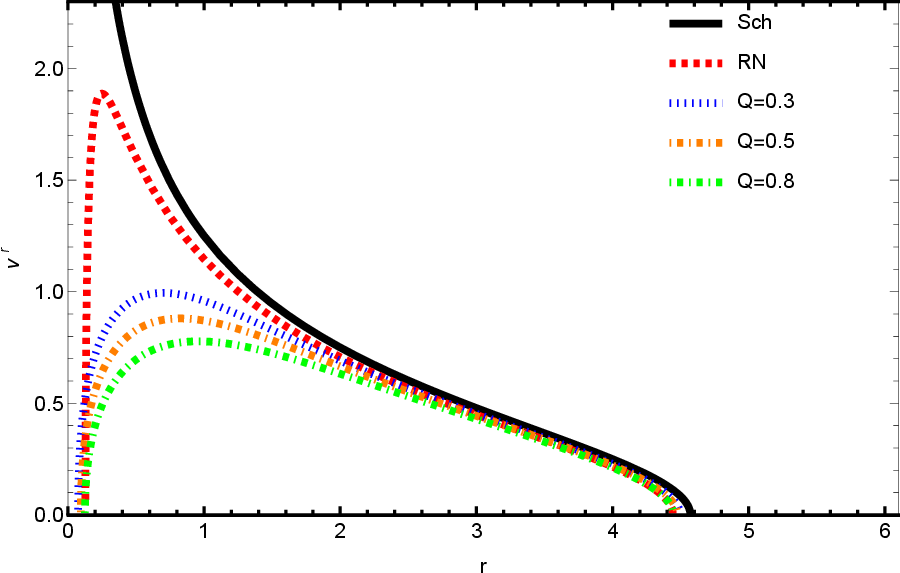}}
\,\,\,
\subfloat[\label{FigVrb} $Q$ = 0.3]{\includegraphics[width=0.475\textwidth]{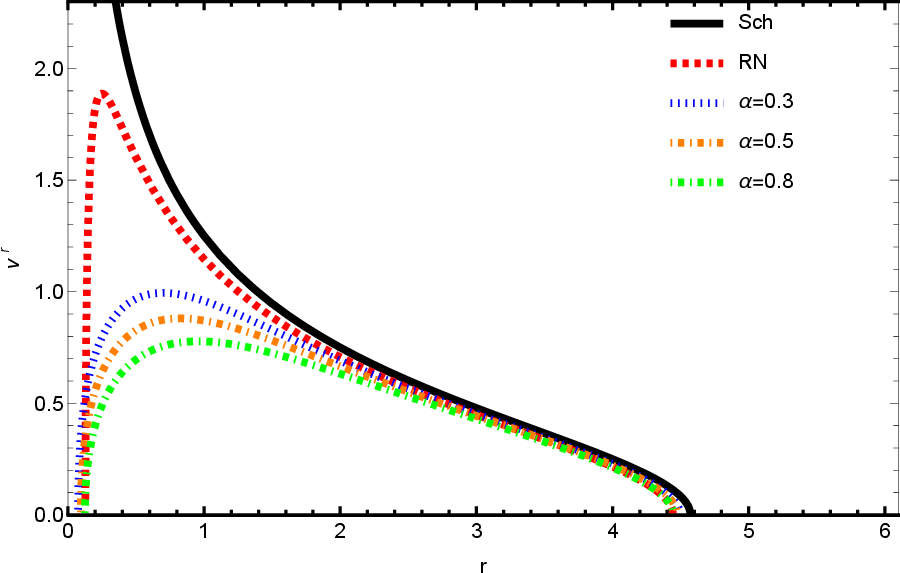}}
\caption{\label{FigVr}\small{\emph{The plot of radial velocity of the charged $4D$ EGB black hole versus r for different values of ${\alpha}$ and Q. The black solid and the red dashed lines are the Schwarzschild and Reissner–Nordström black hole solutions in GR.}}}
\end{figure}

Now inserting $p = \omega \rho \,$ into the Eq. \eqref{25}, the radial velocity of the perfect fluid is as follows
\begin{equation}\label{26}
v^r = \sqrt {\left( {\frac{{{A_4}}}{{\omega  + 1}}} \right)\left( {1 + \frac{{{r^2}}}{{2\alpha }}\left[ {1 - \sqrt {1 + 4\alpha \left( {\frac{{2M}}{{{r^3}}} - \frac{{{Q^2}}}{{{r^4}}}} \right)} } \right]} \right)} \,.
\end{equation}
The graphs of $v^{r}$ against $r$ for the charged $4D$ EGB black hole in comparison to the Schwarzschild and Reissner–Nordström  black holes are shown in Figure \ref{FigVr}. As we can see from Fig. \ref{FigVr}, the fluid with the radial element of its four-velocity matching with each curve (related with each values of $\alpha$ and $Q$) starts to flow from rest at large $r$ towards the charged $4D$ EGB black hole. It then moves toward the charged $4D$ EGB black hole, in order to return to the rest state. Additionally, as we can see in Fig. \ref{FigVr}, a decrease in the parameters $\alpha$ and $Q$ results in an increase in the value of $v^{r}$ at a distance from the source.

\begin{figure}[htb]
\centering
\subfloat[\label{Figrhoa} $\alpha$ = 0.3]{\includegraphics[width=0.475\textwidth]{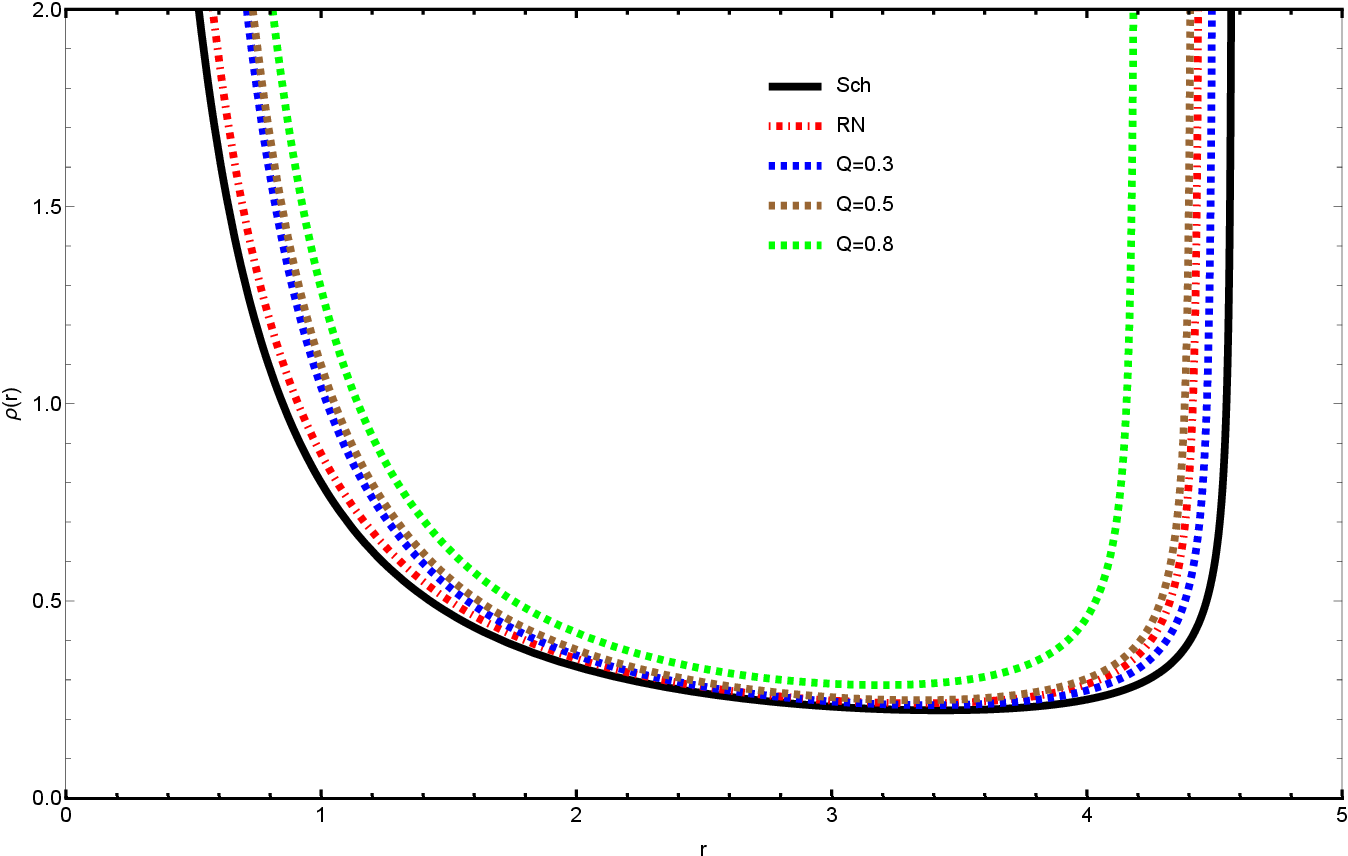}}
\,\,\,
\subfloat[\label{Figrhob} $Q$ =0.3]{\includegraphics[width=0.475\textwidth]{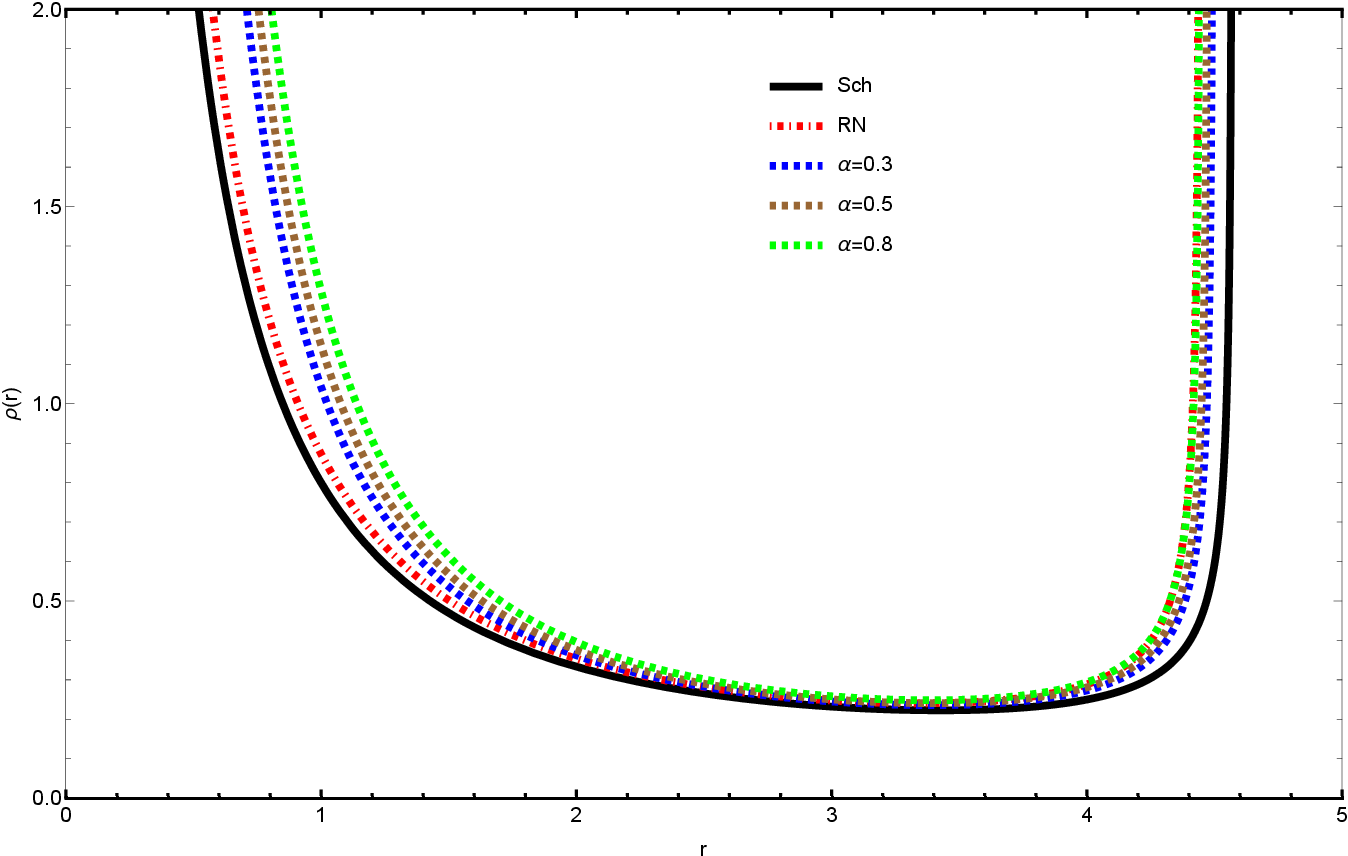}}
\caption{\label{Figrho}\small{The plot of ${\rho}$ of the charged $4D$ EGB black hole versus r for different values of ${\alpha}$ and Q. The black solid and the red dashed lines are the Schwarzschild and Reissner–Nordström black hole solutions in GR.}}
\end{figure}

By putting the Eq. \eqref{26} into the Eq. \eqref{23}, we reach at the specific energy density of the accreting fluid as follows
\begin{equation}\label{27}
{\rho} = \frac{{{A_3}\left( {\omega  + 1} \right)}}{{{r^2}\sqrt {A_4^2 - \left( {1 + \frac{{{r^2}}}{{2\alpha }}\left[ {1 - \sqrt {1 + 4\alpha \left( {\frac{{2M}}{{{r^3}}} - \frac{{{Q^2}}}{{{r^4}}}} \right)} } \right]} \right){{\left( {\omega  + 1} \right)}^2}} }}\,.
\end{equation}
The graph of $\rho$ vs $r$ for the charged $4D$ EGB black hole is shown in Figure \ref{Figrho} for different values of EGB parameter $\alpha$ and charge $Q$. It is compared to the Schwarzschild and Reissner–Nordström black holes.

\subsection{Mass evolution}

Both for dark compact objects and black holes, the central source mass is a dynamic quantity over time. For instance, the accretion process causes them to accumulate more mass by accreting surrounding matter onto them. The charged $4D$ EGB black hole's mass change measure or accretion rate can be found using  $\frac{{dM}}{{dt}} =  - \int {T_t^rds}\,$ where the black holes's surface elements is $ds=\left(\sqrt{-g}\right)d\theta d\varphi$ and $T^{r}_{t}=(p+\rho)v_{t}v^{r}$. As a result, $\dot{M}$, accretion rate can be found as 
\begin{equation}\label{28}
\dot M =  - 4\pi {r^2}v\left( {p + \rho } \right)\sqrt {{v^2} + 1 + \frac{{{r^2}}}{{2\alpha }}\left[ {1 \pm \sqrt {1 + 4\alpha \left( {\frac{{2M}}{{{r^3}}} - \frac{{{Q^2}}}{{{r^4}}}} \right)} } \right]} \,.
\end{equation}

Redefining ${A_0} =  - {A_1}{A_2}\,$ and ${A_2} = {M^2}\left( {{p_\infty } + {\rho _\infty }} \right)\sqrt {F\left( {{r_\infty }} \right)} \,$, the Eq \eqref{28} becomes
\begin{equation}\label{29}
\dot M = 4\pi {A_1}{M^2}\left( {{p_\infty } + {\rho _\infty }} \right)\sqrt {F\left( {{r_\infty }} \right)} \,.
\end{equation}

By scrutinizing in the Eq \eqref{29}, we find the equation between the initial mass and the mass at the arbitrary time $t$ as follows
\begin{equation}\label{30}
{M_f} \equiv \frac{{{M_i}}}{{1 - \frac{t}{{{t_{cr}}}}}}\,,
\end{equation}
where ${t_{cr}}\,$ is the critical time defined as

$${t_{cr}} = \frac{1}{{4\pi {A_1}{M_i}\left( {p + \rho } \right)\sqrt {F\left( {{r_\infty }} \right)} }}$$,

where $M_i$ is considered as initial mass. At this critical point, the mass of the charged $4D$ EGB black hole approaches infinity in a finite time.

\subsection{Critical accretion}\label{BAorCA}

Due to the central source's gravitational field, the fluid at rest in a far distance from the source (charged $4D$ EGB black hole) starts to flow inward throughout the accretion phase and keeps accelerating. The fluid reaches a sonic (critical) point during its inward flow motion toward the source, where its four velocities coincide with the local speed of sound $c_{s}$. Supersonic velocities are reached by the inward flow accelerated motion from this critical point to the core source. To find the critical point, there must be a radial velocity gradient. 

By deriving in terms of $r$ from Eqs. \eqref{24} and \eqref{25}, we obtain
\begin{equation}\label{31}
\frac{{{\rho ^\prime }}}{\rho } + \frac{{{v^\prime }}}{v} + \frac{2}{r} = 0\,
\end{equation}

and
\begin{equation}\label{32}
\frac{{{\rho ^\prime }}}{\rho }\left[ {\frac{{d\ln \left( {p + \rho } \right)}}{{d\ln \left( \rho  \right)}} - 1} \right] + \frac{{v{v^\prime }}}{{{v^2} + 1 + \frac{{{r^2}}}{{2\alpha }}\left[ {1 - \sqrt {1 + 4\alpha \left( {\frac{{2M}}{{{r^3}}} - \frac{{{Q^2}}}{{{r^4}}}} \right)} } \right]}} + \frac{{\frac{{{\rm{ - }}2M\alpha  + {r^3}\left( { - 1 + \sqrt {1 - \frac{{4\left( {{Q^2} - 2Mr} \right)\alpha }}{{{r^4}}}} } \right)}}{{{r^2}\alpha \sqrt {1 - \frac{{4\left( {{Q^2} - 2Mr} \right)\alpha }}{{{r^4}}}} }}}}{{2\left( {{v^2} + 1 + \frac{{{r^2}}}{{2\alpha }}\left[ {1 - \sqrt {1 + 4\alpha \left( {\frac{{2M}}{{{r^3}}} - \frac{{{Q^2}}}{{{r^4}}}} \right)} } \right]} \right)}} = 0\,.
\end{equation}

By redefinition we have
\begin{equation}\label{33}
\frac{{d\ln \left( v \right)}}{{d\ln \left( r \right)}} = \frac{{{\varepsilon _1}}}{{{\varepsilon _2}}} \Rightarrow \left\{ {\begin{array}{*{20}{c}}
{{\varepsilon _1} = \frac{{\frac{{{\rm{ - }}2M\alpha  + {r^3}\left( { - 1 + \sqrt {1 - \frac{{4\left( {{Q^2} - 2Mr} \right)\alpha }}{{{r^4}}}} } \right)}}{{r\alpha \sqrt {1 - \frac{{4\left( {{Q^2} - 2Mr} \right)\alpha }}{{{r^4}}}} }}}}{{2\left( {{v^2} + 1 + \frac{{{r^2}}}{{2\alpha }}\left[ {1 - \sqrt {1 + 4\alpha \left( {\frac{{2M}}{{{r^3}}} - \frac{{{Q^2}}}{{{r^4}}}} \right)} } \right]} \right)}} - 2{V^2}}\\
\\
{{\varepsilon _2} = {V^2} - \frac{{{v^2}}}{{{v^2} + 1 + \frac{{{r^2}}}{{2\alpha }}\left[ {1 - \sqrt {1 + 4\alpha \left( {\frac{{2M}}{{{r^3}}} - \frac{{{Q^2}}}{{{r^4}}}} \right)} } \right]}}}
\end{array}} \right.\,,
\end{equation}
in which
\begin{equation}\label{34}
{V^2} = \frac{{d\ln \left( {p + \rho } \right)}}{{d\ln \left( \rho  \right)}} - 1\,.
\end{equation}
To find the critical point, the condition ${\varepsilon _1} = {\varepsilon _2} = 0\,$ must be satisfied \cite{Armitage_2020}. The existence of this condition leads to two consequences; the first consequence is the critical speed as
\begin{equation}\label{35}
V_c^2 = \frac{{{\rm{ - }}2Mr\alpha  + {r^4}\left( {{\rm{ - }}1 + \sqrt {1 - \frac{{4({Q^2} - 2Mr)\alpha }}{{{r^4}}}} } \right)}}{{ - 8{Q^2}\alpha {\rm{ - }}18Mr\alpha  + 4{r^2}\alpha \sqrt {1 - \frac{{4({Q^2} - 2Mr)\alpha }}{{{r^4}}}}  + 3{r^4}\left( {{\rm{ - }}1 + \sqrt {1 - \frac{{4({Q^2} - 2Mr)\alpha }}{{{r^4}}}} } \right)}}\,.
\end{equation}

The range of the critical radius is determined by the positivity of its denominator, as shown by the following inequality
\begin{equation}\label{36}
\frac{{ - 8{Q^2}\alpha {\rm{ - }}18Mr\alpha  + 4{r^2}\alpha \sqrt {1 - \frac{{4({Q^2} - 2Mr)\alpha }}{{{r^4}}}}  + 3{r^4}\left( {{\rm{ - }}1 + \sqrt {1 - \frac{{4({Q^2} - 2Mr)\alpha }}{{{r^4}}}} } \right)}}{{{r^2}\alpha \sqrt {1 - \frac{{4({Q^2} - 2Mr)\alpha }}{{{r^4}}}} }} > 0\,.
\end{equation}
Furthermore, as the second consequence the criterion for critical points provides us
\begin{equation}\label{37}
(v^{r}_c)^2 = \frac{1}{4} r \left(\frac{r \left(1-\sqrt{\frac{8 \alpha  M}{r^3}-\frac{4 \alpha  Q^2}{r^4}+1}\right)}{\alpha }-\frac{r^2 \left(\frac{16 \alpha  Q^2}{r^5}-\frac{24 \alpha  M}{r^4}\right)}{4 \alpha  \sqrt{\frac{8 \alpha  M}{r^3}-\frac{4 \alpha  Q^2}{r^4}+1}}\right).
\end{equation}

Finally, the local speed of sound for accretion of charged $4D$ EGB black hole is
\begin{equation}\label{38}
c_s^2 =  - 1 + {A_4}\sqrt {(v^{r}){^2} + 1 + \frac{{{r^2}}}{{2\alpha }}\left[ {1{\rm{ - }}\sqrt {1 + 4\alpha \left( {\frac{{2M}}{{{r^3}}} - \frac{{{Q^2}}}{{{r^4}}}} \right)} } \right]} \,.
\end{equation}

\begin{figure}[htb]
\centering
\subfloat[\label{Figvc1} $\alpha$ = 0.3]{\includegraphics[width=0.475\textwidth]{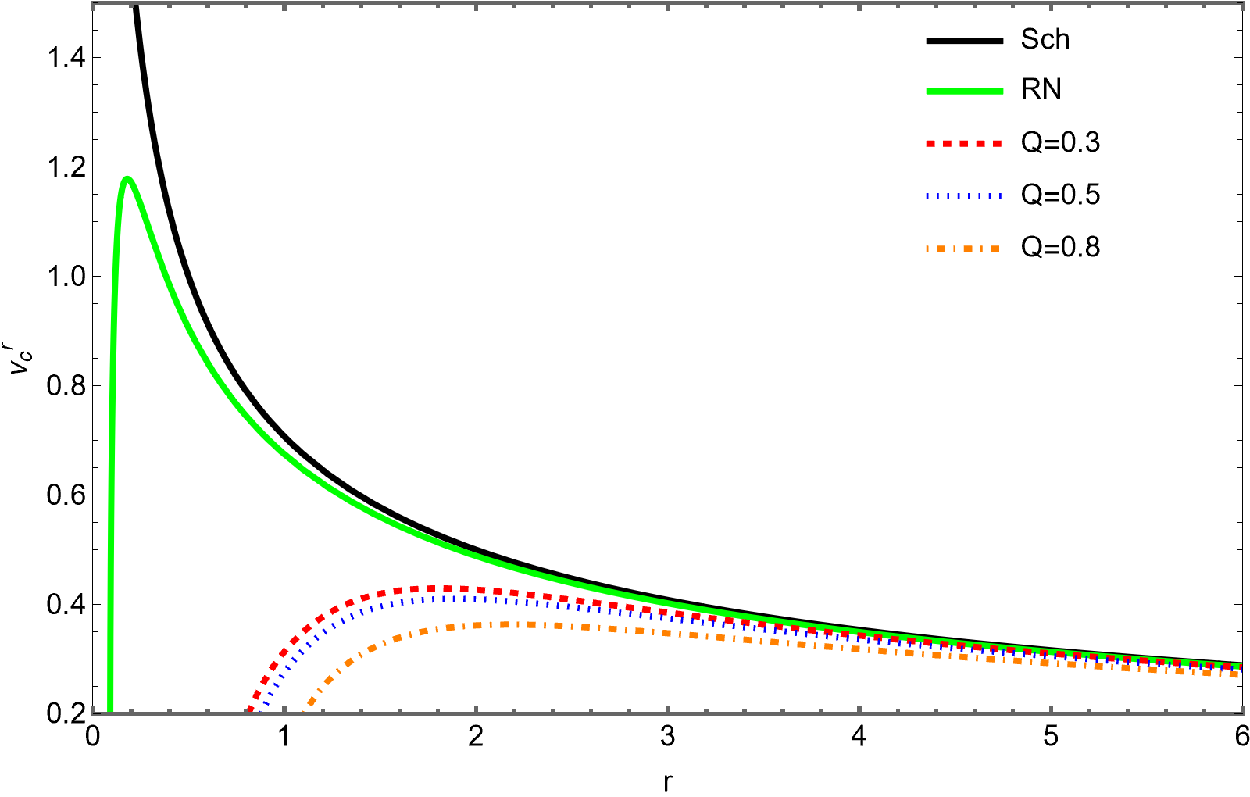}}
\,\,\,
\subfloat[\label{Figvc2} $Q$ =0.3]{\includegraphics[width=0.475\textwidth]{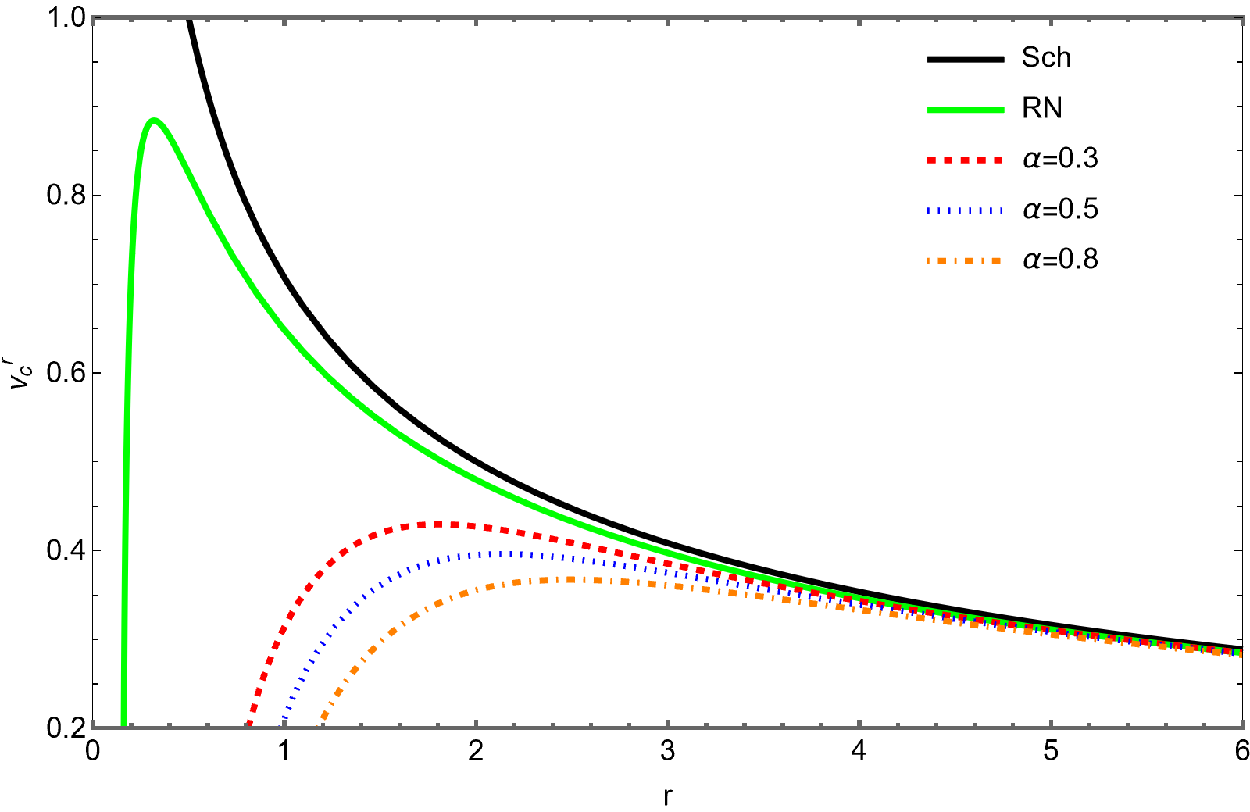}}
\caption{\label{Figvc}\small{The plot of $v^{r}_c$ of the charged $4D$ EGB black hole versus r for different values of ${\alpha}$ and Q. The black solid and the red dashed lines are the Schwarzschild and Reissner–Nordström black hole solutions in GR.}}
\end{figure}

Figure \ref{Figvc}  is the graph of the critical velocity 
versus $r$ for the charged $4D$ EGB black hole in
comparison with Schwarzschild and Reissner–Nordström black holes in GR.

\section{Summary and conclusions}\label{SaC}

The $4D$ EGB gravity theory proposes Gauss-Bonnet curvature corrections to the Einstein term, with a coupling constant proportional to $\alpha/(D - 4)$ in the limit $D \rightarrow 4$.
This equation demonstrates that $4D$ EGB gravity can significantly impact gravitational dynamics in the limit $D \rightarrow 4$. Several recent studies have found that the static, spherically symmetric black hole solution remains consistent across different theories. This paper examines the effects of the EGB parameter and electric charge $Q$ on particles' geodesics in the charged $4D$ EGB black hole spacetime. We have studied the motion of test particles in the environment surrounding the black hole spacetime geometry and also the radiation properties of the accretion disk around the charged black hole in EGB thoery. As, the accretion disk radiation can be influenced by the charged $4D$ EGB black hole parameters, the theoretical study of accretion disk radiation around a charged black hole in EGB theory is considered to be a powerful test of its useful properties and astrophysical observations. We compared our results to those of Reissner–Nordström ($\alpha \rightarrow 0$) and Schwarzschild black holes ($Q \rightarrow 0$ and $\alpha \rightarrow 0$ ). The impacts of the Gauss-Bonnet coupling constant $\alpha$ and charge on physical quantities of accretion process have been clearly demonstrated. 

 First, we investigated the effective potential for the test particles. It was discovered that increasing the value of the EGB parameter $\alpha$ increase the effective potential of the  particle moving in the spacetime of the charged $4D$ EGB black hole for a fixed value of $Q$. On the other hand by fixing $\alpha$, increasing $Q$ amplifies the effective
potential of a test particle in a charged 4D EGB black hole spacetime. Furthermore, we established that the parameter $\alpha$ of the EGB setup decreases the specific energy, angular momentum and angular velocity of the test particle, also this situation stands for amplifying the values of charge $Q$. We also demonstrated that the EGB parameter $\alpha$ and charge $Q$ both diminish the ISCO radius of test particles. We observed, however, that the ISCO radius of the test particle associated with the charged $4D$ EGB black hole is smaller than that of the Schwarzschild and Reissner–Nordström black holes in GR. Further, we investigated the radiative efficiency and presented the results in Table \ref{Table1}. It was proven that the radiative efficiency grows with rising the EGB parameter $\alpha$, and again, it increases as the value of the electric charge $Q$ becomes more.

In EGB theory, the accretion disk around a black hole is a key source of information about its surrounding spacetime geometry and its nature. So in the next step, we have studied the impact of the EGB parameter on the accretion disk's radiation properties, including flux, temperature and differential luminosities, to gain insight into the charged $4D$ black hole's unique properties in EGB theory. We discovered that as the EGB  parameter $\alpha$ increases, the curves of accretion disk radiation quantities shift upwards, toward greater values. The study includes not only circular orbits, their features and accretion disk properties but also epicyclic frequencies. As we explored, vertical epicyclic frequency decreases monotonically with $r$ and has no extremes. But the radial epicyclic frequency always has a maximum, and the EGB parameter $\alpha$ has a significant impact on it. We found that the dependence on this parameter for radial epicyclic frequency is substantial close to the black hole, but weaker away from the center mass. Increasing the EGB parameter $\alpha$ pushes the maximum of radial epicyclic frequency to smaller radii. Our findings indicate that $\Omega_{r}<\Omega_{\theta}$.

Moreover, we consider spherically symmetric accretion on the equatorial plane. We suppose that the accreting matter is a perfect fluid that flows onto the charged $4D$ EGB black hole. The radial component of the four-velocity and the accreting fluid's energy density decrease when the EGB parameter $\alpha$ increases close to the source, but this is not for the case farther away from the source.

\end{document}